\documentclass[conference]{IEEEtran}

\ifCLASSINFOpdf
\else
\fi

\usepackage{cite}
\usepackage{amsmath,amssymb,amsfonts}
\usepackage{graphicx}
\usepackage{listings}
\usepackage{textcomp}
\usepackage{xcolor}
\def\BibTeX{{\rm B\kern-.05em{\sc i\kern-.025em b}\kern-.08em
    T\kern-.1667em\lower.7ex\hbox{E}\kern-.125emX}}

\usepackage{tikz}
\usepackage{amsmath}
\usepackage{filecontents}
\usepackage{algorithm}
\usepackage{algpseudocode}
\usepackage{xspace}
\usepackage{hyperref}
\usepackage[table]{xcolor}
\usepackage[most]{tcolorbox}
\usepackage{etoolbox}
\usepackage[normalem]{ulem}
\usepackage[nolist]{acronym}

\newcounter{challengecounter}
\renewcommand{\thechallengecounter}{Challenge~\arabic{challengecounter}}

\newtcolorbox{challengebox}[1][]{
  colback=gray!10,
  colframe=gray!90,
  boxrule=0.5pt,
  arc=2mm,
  left=1mm,
  right=1mm,
  top=1mm,
  bottom=1mm,
  fonttitle=\bfseries,
  title=\stepcounter{challengecounter}\thechallengecounter: #1
}

\newcommand{\ufuzzNoSpace}{\fuzzilicon}
\newcommand{\ufuzz}{\ufuzzNoSpace\xspace}
\newcommand{\fuzzilicon}{Fuzzilicon}

\newcommand{\ourname}{\fuzzilicon\xspace}

\newcommand{\ucodeNoSpace}{$\mu$code}
\newcommand{\ucode}{\ucodeNoSpace\xspace}
\newcommand{\uopNoSpace}{$\mu$operation}
\newcommand{\uop}{\uopNoSpace\xspace}
\newcommand{\uopsNoSpace}{$\mu$operations}
\newcommand{\uops}{\uopsNoSpace\xspace}

\newcommand{\uarchNoSpace}{microarchitectural}
\newcommand{\uarch}{\uarchNoSpace\xspace}

\newcommand{\fagentNoSpace}{\textit{Fuzzer Agent}}
\newcommand{\fagent}{\fagentNoSpace\xspace}

\newcommand{\fcontrollerNoSpace}{\textit{Fuzzer Controller}}
\newcommand{\fcontroller}{\fcontrollerNoSpace\xspace}

\newcommand{\fwatchdogNoSpace}{\textit{Watchdog}}
\newcommand{\fwatchdog}{\fwatchdogNoSpace\xspace}

\newcommand{\riscv}{\textit{RISC\mbox{-}V}}
\newcommand{\xx}{\textit{x86}}

\newcommand{\artifactDOI}{\url{https://doi.org/10.5281/zenodo.17012971}\xspace}
\newcommand{\github}{\url{https://github.com/0xCCF4/ufuzz}\xspace}

\newcommand{\totalcovp}{$16.27$\%\xspace}
\newcommand{\totalcovpoint}{$2,867$\xspace}
\newcommand{\possiblecovpoint}{$17,624$\xspace}
\newcommand{\overheadimprove}{$31\times$\xspace}
\newcommand{\findingcnt}{$5$\xspace}

\newcounter{challengeNo}

\newcounter{subChallengeNo}[challengeNo]

\DeclareRobustCommand{\challengeRef}[1]{\hyperref[#1]{C\ref*{#1}}}

\newcommand{\CiteAllFuzzers}{\cite{rfuzz,difuzzrtl,osiris,epex,silifuzz,thehuzz,Bruns2022,hypfuzz,psofuzz,mabfuzz,morfuzz,socfuzz,processorFuzz,surgefuzz,stresstest,chatfuzz,cascade,RISCVuzz,FuzzWiz,wu2025genhuzz,wu2025hfl,borkar2024whisperfuzz}}
\newcommand{\CitePreSiliconFuzzers}{\cite{rfuzz,difuzzrtl,thehuzz,Bruns2022,hypfuzz,psofuzz,mabfuzz,morfuzz,surgefuzz,wu2025genhuzz,wu2025hfl,borkar2024whisperfuzz}}

\newcommand{\WRMSR}{\texttt{WRMSR}}

\definecolor{bg}{RGB}{250,250,250}
\definecolor{graynum}{gray}{0.45}

\newcommand{\circleicon}[1]{%
    \tikz[baseline=(char.base)] \node[draw, circle, fill=white, text=black, inner sep=0.8pt] (char) {\scriptsize #1};%
}

\usepackage{listings}
\usepackage{xcolor}

\lstdefinestyle{asmstyle}{
    language=[x86masm]Assembler,   %
    basicstyle=\ttfamily\small,   %
    keywordstyle=\color{blue},    %
    commentstyle=\color{green!50!black}, %
    stringstyle=\color{red},      %
    numbers=left,                 %
    numberstyle=\tiny\color{gray},%
    stepnumber=1,                 %
    numbersep=10pt,               %
    backgroundcolor=\color{gray!10}, %
    frame=single,                 %
    tabsize=4,                    %
    captionpos=b,                 %
    breaklines=true,              %
    breakatwhitespace=true,       %
    showspaces=false,             %
    showtabs=false,               %
    showstringspaces=false,       %
}

\lstdefinestyle{fancyasm}{
    language=[x86masm]Assembler,       %
    basicstyle=\ttfamily\small,       %
    keywordstyle=\color{violet}\bfseries, %
    commentstyle=\color{green!50!black}\itshape, %
    stringstyle=\color{red},          %
    numbers=in,                     %
    numberstyle=\tiny\color{gray},    %
    stepnumber=1,                     %
    numbersep=10pt,                   %
    backgroundcolor=\color{gray!5},   %
    rulecolor=\color{black},          %
    frame=shadowbox,                  %
    tabsize=4,                        %
    captionpos=b,                     %
    breaklines=true,                  %
    breakatwhitespace=true,           %
    showspaces=false,                 %
    showtabs=false,                   %
    showstringspaces=false,           %
    morekeywords={li, sw, lb, csrr, csrc, csrs, addi}, %
}

\usepackage[available]{ndssbadges}

\begin{document}

\title{{\ourname}: \\ A Post-Silicon Microcode-Guided x86 CPU Fuzzer}

\IEEEoverridecommandlockouts
\makeatletter\def\@IEEEpubidpullup{6.5\baselineskip}\makeatother
\IEEEpubid{\parbox{\columnwidth}{
        Network and Distributed System Security (NDSS) Symposium 2026 \\
        23 - 27 February 2026 , San Diego, CA, USA\\
        ISBN 979-8-9919276-8-0 \\
        https://dx.doi.org/10.14722/ndss.2026.231486\\
        www.ndss-symposium.org
}
\hspace{\columnsep}\makebox[\columnwidth]{}}

\author{
\IEEEauthorblockN{Johannes Lenzen}
\IEEEauthorblockA{Technical University of Darmstadt\\johannes.lenzen@stud.tu-darmstadt.de}
\\
\IEEEauthorblockN{Lichao Wu}
\IEEEauthorblockA{Technical University of Darmstadt\\lichao.wu@trust.tu-darmstadt.de}
\and
\IEEEauthorblockN{Mohamadreza Rostami}
\IEEEauthorblockA{Technical University of Darmstadt\\mohamadreza.rostami@trust.tu-darmstadt.de}
\\
\IEEEauthorblockN{Ahmad-Reza Sadeghi}
\IEEEauthorblockA{Technical University of Darmstadt\\ahmad.sadeghi@trust.tu-darmstadt.de}
}
\maketitle

\pagestyle{plain} %
\setcounter{page}{1} 

\begin{acronym}[CRBUS] %
  \acro{API}[API]{Application Programming Interface}
  \acro{BIOS}[BIOS]{Basic Input/Output System}
  \acro{CISC}[CISC]{Complex Instruction Set Computer}
  \acro{CPU}[CPU]{Central Processing Unit}
  \acro{CRBUS}[CRBUS]{Control Register Bus}
  \acro{DUT}[DUT]{device under test}
  \acro{FPGA}[FPGA]{Field Programmable Gate Array}
  \acro{GDT}[GDT]{Global Descriptor Table}
  \acro{GPIO}[GPIO]{General Purpose Input/Output}
  \acro{IP}[IP]{Instruction Pointer}
  \acro{ISA}[ISA]{Instruction Set Architecture}
  \acro{JTAG}[JTAG]{Joint Test Action Group}
  \acro{LLM}[LLM]{Large Language Model}
  \acro{OEM}[OEM]{Original Equipment Manufacturer}
  \acro{OS}[OS]{Operating System}
  \acro{PML}[PML]{Page Map Level}
  \acro{RAM}[RAM]{Random Access Memory}
  \acro{REST}[REST]{Representational State Transfer}
  \acro{ROM}[ROM]{Read-Only Memory}
  \acro{RTL}[RTL]{Register Transfer Level}
  \acro{SMM}[SMM]{System Management Mode}
  \acro{TSS}[TSS]{Task State Segment}
  \acro{UDP}[UDP]{User Datagram Protocol}
  \acro{UEFI}[UEFI]{Unified Extensible Firmware Interface}
  \acro{USB}[USB]{Universal Serial Bus}
  \acro{VM}[VM]{Virtual Machine}
  \acro{VMCS}[VMCS]{Virtual Machine Control Structure}
\end{acronym}

\begin{abstract}
Modern \acp{CPU} are black boxes, proprietary, and increasingly characterized by sophisticated microarchitectural flaws that evade traditional analysis. While some of these critical vulnerabilities have been uncovered through cumbersome manual effort, building an automated and systematic vulnerability detection framework for real-world post-silicon processors remains a challenge.

In this paper, we present \ourname, the first post-silicon fuzzing framework for real-world x86 \acp{CPU} that brings deep introspection into the microcode and microarchitectural layers. \ourname automates the discovery of vulnerabilities that were previously only detectable through extensive manual reverse engineering, and bridges the visibility gap by introducing microcode-level instrumentation. At the core of \ourname is a novel technique for extracting feedback directly from the processor's microarchitecture, enabled by reverse-engineering \textit{Intel}'s proprietary microcode update interface. We develop a minimally intrusive instrumentation method and integrate it with a hypervisor-based fuzzing harness to enable precise, feedback-guided input generation, without access to \ac{RTL} or vendor support.

Applied to \textit{Intel}'s \textit{Goldmont} microarchitecture, \ourname introduces \findingcnt significant findings, including two previously unknown microcode-level speculative-execution vulnerabilities. Besides, the \ourname framework automatically rediscover the $\mu$Spectre class of vulnerabilities, which were detected manually in the previous work. \ourname reduces coverage collection overhead by up to \overheadimprove compared to baseline techniques and achieves \totalcovp unique microcode coverage of hookable locations, the first empirical baseline of its kind. As a practical, coverage-guided, and scalable approach to post-silicon fuzzing, \ourname establishes a new foundation to automate the discovery of complex \ac{CPU} vulnerabilities.

\end{abstract}

\IEEEpeerreviewmaketitle

\section{Introduction}

Computation and information processing form the backbone of modern society, and their security fundamentally depends on the trustworthiness of the underlying hardware. At the core of secure computing are \acp{CPU}, which are expected to faithfully implement their \acp{ISA} and enforce strict isolation between processes. However, this assumption has been increasingly challenged by the discovery of critical architectural and \uarch-level vulnerabilities~\cite{zenbleed,reptar,lipp2018meltdown,kocher2020spectre}. These attacks demonstrate that flaws in CPU \uarch can be exploited to leak data, bypass protections, or undermine system integrity, even for secure and well-written software~\cite{wiebing_inspectre_2024}.
Indeed, modern processors, particularly in the x86 family, are highly complex, with layers of undocumented behavior implemented in proprietary \ucode~\cite{Koppe2017reverse}. As designs become increasingly complex and opaque, the risk of hardware-level security flaws continues to grow~\cite{nist8517}.

To detect hardware-level vulnerabilities, researchers have traditionally relied on techniques such as formal verification~\cite{sarangi2006phoenix,deutschbein2018mining,wile2005comprehensive,dessouky2019hardfails,clarke2011model}, runtime detection~\cite{wagner2007engineering,hicks2015specs}, information flow tracking~\cite{li2011caisson,li2014sapper,zhang2015hardware}, and hardware fuzzing~\cite{kande2022thehuzz,rostami2024fuzzerfly}. 
Among them, hardware fuzzing has emerged as a promising approach due to its scalability and adaptability to various designs~\CiteAllFuzzers. 
Hardware fuzzing has evolved into two distinct approaches: pre-silicon fuzzing, which targets Register-Transfer Level (\ac{RTL}) models during hardware development, and post-silicon fuzzing, which evaluates manufactured processors under real execution conditions~\CiteAllFuzzers. 
While pre-silicon fuzzing is widely studied in literature thanks to the deep observability and fine-grained instrumentation within the \ac{RTL} model~\CitePreSiliconFuzzers, post-silicon fuzzing is rarely touched.
The reason is straightforward: post-silicon fuzzers commonly target black-box or proprietary \acp{CPU} (e.g., from Intel and AMD) with visibility limited to architectural registers or crash symptoms~\cite{silifuzz}. 
Even worse, the internal \uarch state and \ucode-level behavior, where many subtle bugs manifest~\cite{zenbleed,uSpectre,reptar}, are largely inaccessible and undocumented. Existing hardware feedback mechanisms, such as performance counters or architectural registers, offer only coarse-grained or indirect insight. The lack of transparency and informative feedback prevents the evaluator from finding unexpected behaviors and tracing corresponding root causes.

\noindent 
\textbf{Our Contribution.} 
In this work, we present \ufuzz, the first post-silicon fuzzer for proprietary x86 \acp{CPU} with gray-box visibility. \ufuzz introduces a novel internal microarchitectural feedback channel to guide test generation. By running the \ac{CPU} in a Red-unlocked mode~\cite{undocumented-instructions} and leveraging undocumented debugging and instrumentation capabilities in Intel processors~\cite{chip-red-pill}, we gain access to the \ucode engine interface. We re-purpose this interface, which is typically used to deploy \ucode patches, as a programmable introspection layer, inserting lightweight instrumentation directly into the processor. Through careful reverse engineering, we construct \ucode patches that instrument internal \ucode execution paths. This turns a proprietary \ac{CPU} into a gray box, enabling observation of internal execution states (e.g., \ucode path transitions) at runtime, without \ac{RTL} access or specialized hardware.
To ensure safe and deterministic execution of fuzzing workloads on the target \ac{CPU}, we build a bare-metal, hypervisor-based fuzzing framework that isolates the \ac{DUT}, controls its environment, and continuously monitors execution. We further introduce a serialization oracle that synthesizes semantically equivalent variants of instruction sequences, improving fuzzing reproducibility and enabling reliable detection of vulnerabilities and divergences across microarchitectural implementations.
Together, these capabilities enable feedback-driven fuzzing of real, post-silicon x86 processors with microarchitectural visibility, uncovering rare execution paths and vulnerabilities.

\noindent Our contributions are listed as follows:

\begin{itemize}
  \item We introduce \ufuzz, the first post-silicon x86 \ac{CPU} fuzzer that leverages internal \uarch feedback by injecting runtime instrumentation using \ucode patches, enabling introspection on proprietary silicon.
  
  \item We introduce a new \ucode coverage feedback for post-silicon \ac{CPU} fuzzing, enabling visibility into CPU's internal execution at the granularity of \ucode operations.
  
  \item We design an optimized \ucode instrumentation strategy that reduces patching overhead by 31 times compared to the baseline instrumentation~\cite{CustomProcessingUnit}.
  
  \item We demonstrate a novel \ucode-level speculative execution fuzzing use case, enabling \ufuzz to detect \ucode-level speculative vulnerabilities and reveal undocumented leakage paths.

  \item On \textit{Intel N3350 (Goldmont)} \ac{CPU}, \ufuzz uncovered \findingcnt significant findings, including two previously unknown \ucode-level speculative-execution vulnerabilities.
   
  \item The \ufuzz framework automatically rediscover the $\mu$Spectre class of vulnerabilities~\cite{uSpectre}, which previously was detected manually.
  
  \item We extend prior reverse engineering of Intel \textit{Goldmont}'s \ucode patching infrastructure to enable, for the first time, custom instrumentation at arbitrary micro-op entry points.
  
  \item We design a low-overhead and bare metal-hypervisor for \ac{CPU} fuzzing that safely isolates arbitrary x86 programs and preserves determinism and system stability even under malformed instruction sequences.

  \item We introduce a serialization oracle that generates semantically equivalent instruction sequences, enabling robust cross-platform divergence detection without any ground-truth architectural oracle.
  
\end{itemize}

The complete source code for the \ourname{} framework is open-sourced at \github and permanently publicly available at \artifactDOI. For detailed instructions on setting up and using the framework, please refer to Appendix~\ref{appendix:artifact}.

The rest of this paper is organized as follows. Section~\ref{cha:bg} provides background on key concepts necessary for understanding \ufuzz, including x86 \ucode execution, instruction decoding, \textit{Red-Unlock} mode, and \uarch introspection techniques. Section~\ref{sec:challenges} details the core technical challenges of applying coverage-guided fuzzing to commercial x86 \acp{CPU}. Section~\ref{sec:design} outlines the design of \ufuzz, while Section~\ref{sec:implementation} describes its framework implementation, including \ucode instrumentation and control infrastructure. Section~\ref{sec:evaluation} evaluates \ufuzz's effectiveness in terms of discovered vulnerabilities, coverage, and performance.  Section~\ref{sec:disc} discusses the \ourname with more insights. Section~\ref{sec:related-work} discusses related works. Section~\ref{sec:conclusion} concludes this work.

\section{Background} \label{cha:bg}

\subsection{Microcode and Instruction Decoding} \label{cha:bg:microcode}

Intel x86 instructions range from simple arithmetic operations to highly complex instructions involving system state, memory ordering, or cryptographic operations.
Implementing each \xx\ instruction entirely in hardware would be infeasible due to silicon cost, complexity, and the need for update flexibility~\cite{Koppe2017reverse}. Instead, modern Intel \acp{CPU} use a \ucode engine to decode complex instructions into sequences of simpler \uops, which are then executed by the processor~\cite{microcode-ccc}.
Each x86 instruction could be mapped to one or more \uops. Simple instructions often decode statically into \uops~\cite{Koppe2017reverse}; complex ones invoke the \ucode engine~\cite{Koppe2017reverse}. The \ucode engine expands each \xx\ instruction into a sequence of \uops, scheduled and executed under the structured control-flow primitive known as a \textit{triad}~\cite{undocumented-instructions, CustomProcessingUnit}.
A \textit{triad} contains three \uops and a \textit{sequence word} that controls static branching, ordering, and instruction termination.

\begin{figure*}
  \centering
  \includegraphics[width=0.9\textwidth]{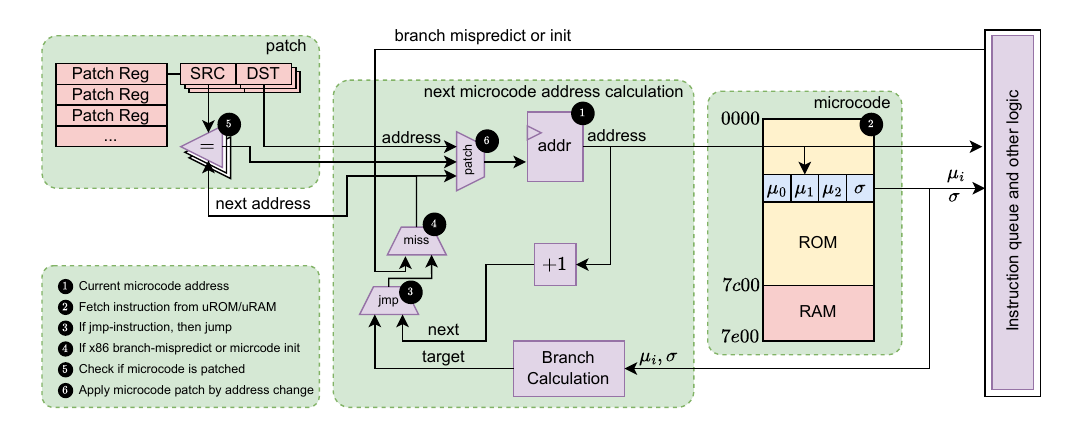}
  \caption{\label{fig:microcode} Simplified view of the Intel \ucode engine~\cite{microcode-ccc}. Red components indicate runtime reconfigurable structures.}
\end{figure*}

Figure~\ref{fig:microcode} presents a high-level view of the \ucode system. The \ac{ROM}, \circleicon{2}, stores factory-installed instruction handlers, whereas the \ucode \ac{RAM} stores runtime patches that manufacturers can apply software patches to even after the \ac{CPU} has shipped. The \ac{CPU}'s \ucode engine first checks the \ac{RAM} for patches before executing the \ac{ROM} default. A set of patch registers stores $(SRC_i \rightarrow DST_i)$ mappings, \circleicon{3} when a \ucode address $SRC_i$ is fetched, execution is redirected to $DST_i$ in \ac{RAM}~\cite{CustomProcessingUnit}. 

\subsection{Red-Unlock Mode and Microarchitectural Access} \label{cha:bg:red-unlock}

Intel \acp{CPU} support multiple debug access levels~\cite{undocumented-instructions, CustomProcessingUnit}: 1) \emph{Green-locked:} default user/OS-visible mode on retail \acp{CPU}, all debugging features are disabled; 2) \emph{Orange-unlocked:} Intended for \acp{OEM}, enables a subset of debug features, and 3) \emph{Red-unlocked:} Engineering mode used internally by Intel; enables unrestricted access to \uarch components.
Specifically, in \textit{Red-unlocked} mode, two undocumented new instructions become available: \texttt{udbgrd} and \texttt{udbgwr}\cite{undocumented-instructions}. These enable reading and writing internal \ac{CPU} memory regions, including the \ucode \ac{RAM}. By writing a sequence of \ucode \textit{triads} and triggering execution at a target address, arbitrary \ucode can be executed. This capability allows bypassing the normal and protected update procedure for \ucode patches, i.e., first signature validation, then \ac{RAM} writing, finally patch register configuration, permits direct manipulation of the currently deployed \ucode update~\cite{CustomProcessingUnit}.

\subsection{Fuzzing} \label{cha:bg:hardware-fuzzing}
Software fuzzing involves generating random and unexpected inputs to explore different execution path and uncover bugs~\cite{afl,libafl,rostami2024fuzzerfly}. Since exhaustive input enumeration is infeasible, fuzzers rely on heuristics, commonly guided by code coverage, to mutate inputs in ways that maximize program exploration~\cite{processorFuzz}. Bugs are typically detected through observable runtime violations, such as crashes, memory access errors, or failed assertions.

Hardware fuzzing adapts these core ideas to the context of processor testing. Instead of application-level inputs, hardware fuzzers generate short programs, i.e., instruction sequences, that are executed directly on the target \ac{CPU}. As in software fuzzing, heuristics are required to guide test generation. These may include \ac{RTL} signal toggling in simulation-based setups or \uarch activity in post-silicon environments. However, detecting bugs in hardware poses unique challenges: \acp{CPU} are not considered ``buggy'' because they fault or access invalid memory regions or trigger exceptions. This is often correct behavior under certain conditions. As a result, post-silicon hardware fuzzing typically depends on two main strategies for bug detection: 1) Assertion checking, which is only feasible when \ac{RTL} is available; and 2) Differential testing, which compares execution results across different \acp{CPU} or hardware configurations to identify inconsistencies~\cite{rostami2024fuzzerfly}. 

Further, post-silicon fuzzing is constrained by restricted visibility into the CPU internal state of the processor, making it difficult to detect complex errors. In this work, we address this limitation by introducing custom instrumentation into the \ucode layer. This allows us to collect CPU internal execution data at runtime, enabling a new class of post-silicon fuzzing that observes \uarch behavior.

\section{Breaking the Post-Silicon Barrier}
\label{sec:challenges}

Despite dominating modern computing, commercial x86 \acp{CPU} have not been the target of coverage-guided fuzzing, a proven method for exposing complex software bugs. This gap is driven by four core challenges: \emph{\uarch invisibility}, \emph{absence of bug detection oracle}, \emph{non-deterministic execution}, and \emph{fault containment}. We detail each challenge in the following and explain how \ufuzz overcomes them.

\begin{challengebox}[Microarchitectural Invisibility]
x86 processors are complex and opaque, with undocumented \ucode and speculative behaviors. This lack of visibility severely limits the feedback needed for the fuzzer.
\end{challengebox}

\noindent Coverage-guided fuzzing depends on informative feedback to guide test generation. However, unlike open architectures such as RISC-V, where pre-silicon emulation allows fine-grained instrumentation, x86 processors are closed-source and proprietary; thus, valuable sources of feedback are limited. While prior work~\cite{silifuzz,osiris} has used coarse-grained feedback, such as general-purpose register values or performance counters, these signals reveal only surface-level execution behavior. They offer limited guidance for exploring the deep \uarch behaviors that carry critical hardware vulnerabilities. Consequently, rich and informative feedback is lacking to drive effective fuzzing campaigns for \xx~processors.

\noindent\textbf{Our Solution.} \ufuzz introduces the first \uarch feedback mechanism for post-silicon fuzzing: \emph{\ucode-level coverage}. By instrumenting \uops, \ufuzz collects fine-grained \uarch traces during execution. This coverage metric enables deep exploration of all \ac{CPU}'s \uarch behavior (\uops) space without requiring \ac{RTL} models. We detail the \ucode-level coverage in Section~\ref{sec:impl-microcode}.

However, enabling \ucode instrumentation is not straightforward. First, the \ucode architecture is proprietary and undocumented. As a result, all public knowledge about Intel's \ucode originates from reverse engineering, and is neither complete nor guaranteed to be correct. This makes deploying custom \ucode updates inherently risky, as unsound patches may corrupt the internal state of the \ac{CPU}. To overcome this, \ufuzz minimizes the state changes originating from \ucode instrumentation, described in Section~\ref{sec:impl-microcode}. Second, prior work~\cite{CustomProcessingUnit} records only whether a given \ucode address executed, ignoring execution multiplicity. Consequently, two x86 instructions that invoke the same \uops but with different iteration counts (e.g., once vs. four times) appear indistinguishable, and one may be discarded for not increasing coverage, despite exploring different \uarch behaviors. This coarse signal misses substantial exploration space and prunes valuable inputs. \ufuzz addresses this by counting executions per \ucode handler, revealing loops, and deeply nested paths within a single test. On our target, \textit{Intel Apollo Lake (Celeron, Goldmont) N3350 (CPUID[1].EAX=0x506ca)}, we repurpose 16 internal registers to track 32 \ucode addresses concurrently. Given a $32k$ \ucode address space, this requires $1k$ instrumentation points. However, we developed coverage-scheduling optimizations to keep the overhead practical (Section~\ref{sec:impl-cov-schedule}).

\begin{challengebox}[Absence of Bug Detection Oracle]
Detecting x86 \ac{CPU} bugs is difficult without formal \uarch specifications. \ac{ISA}-level models miss undocumented/speculative behaviors, limiting bug detection.
\end{challengebox}

\noindent Detecting incorrect behaviors in commercial x86 \acp{CPU} is inherently difficult due to the absence of formal \uarch specifications. While \ac{ISA}-level simulators offer partial reference behavior, they fail to capture \uarch effects such as speculative execution, undocumented instructions, or \ucode operations. As a result, it's often unclear whether observed differences reflect genuine vulnerabilities or benign implementation variations. In the absence of a reliable oracle, fuzzers rely on differential detection: input-driven, which requires known outputs (rare in fuzzing), and output-driven, which compares results across different \ac{CPU} implementations. While general, output-driven methods suffer from 1) False Positives (FP), due to architectural differences, and 2) False Negatives (FN), when different \acp{CPU} exhibit the same flawed behavior. A few solutions have been proposed to address this challenge; for instance, Reversi~\cite{wagner2008reversi} proposes generating a reverse instruction for each instruction. However, for complex \xx~instruction, automatic generation of such reverse instructions is challenging~\cite{zenbleed}.

\noindent\textbf{Our Solution.} Inspired by Zenbleed~\cite{zenbleed}, \ufuzz introduces \emph{Serialization Oracles} (Section~\ref{sec:impl-differential-testing}). For each test case \(P\), \ufuzz synthesizes a semantically equivalent variant \(Q\) by inserting serialization fences to suppress speculation and reordering. Divergences between \(P\) and \(Q\) potentially flag \uarch bugs; no reference model is required~\cite{zenbleed}.

\begin{challengebox}[Fault Containment]
Applying custom \ucode patches in the \ac{CPU} is inherently risky: malformed sequences can corrupt architectural state or block forward progress, stalling or terminating the fuzzing campaign.
\end{challengebox}\label{chal:fault-containment}
\noindent Since \ucode documentation is proprietary and our knowledge originates from reverse engineering, deploying custom \ucode sequences carries nontrivial risk. Exercising undocumented behaviors or injecting malformed \ucode can induce hangs, crashes, or persistent/transient lockups requiring a hardware reset. Without proper fault containment, such failures crash the entire fuzzing infrastructure, resulting in lost execution state and requiring manual recovery.

\noindent\textbf{Our Solution.} \ufuzz decouples test execution from control logic using a two-part architecture. The \emph{Fuzzer Agent}, running on the target \ac{CPU}, is responsible for setting up the hypervisor, executing fuzzing test cases, applying \ucode instrumentation, and collecting internal coverage feedback. The \emph{Fuzzer Controller}, on a separate host, handles test case generation and mutation, coverage analysis, feedback-driven scheduling, and vulnerability triage. If the agent crashes or stalls, the controller resets the system and resumes testing. This architecture ensures that crashes in the system under test do not affect the fuzzer's control logic or analytics pipeline. Furthermore, this architecture enables scalable deployment across multiple agents. We detail our solution in Section~\ref{sec:design}.

\begin{challengebox}[Non-deterministic Execution]
Fuzzing depends on reproducible execution to analyze crashes, however stateful instructions, \ac{OS} noise, and \uarch states introduce Non-determinism during \ac{CPU} fuzzing.
\end{challengebox}\label{chal:nondeter}
\noindent For sound evaluation and triage, fuzzing results must be reproducible~\cite{klees2018evaluating}. Achieving reproducibility on post-silicon x86 \acp{CPU} is challenging due to: stateful instructions (e.g., \texttt{RDRAND}, \texttt{RDTSC}, \texttt{RDPMC}), \ac{OS}/system noise, and \uarch state all introduce nondeterminism. Moreover, fuzzing inputs can corrupt global state or leave persistent side effects that contaminate subsequent tests: unconstrained programs may modify control registers, I/O ports, or memory mappings.

\noindent \textbf{Our Solution.} \ufuzz executes each test case in a bare-metal hypervisor environment that provides strong isolation from the host system, described in Section~\ref{sec:design:hypervisor}. It resets memory and \ac{CPU} state between tests, suppresses asynchronous events, and prevents residual side effects. This ensures repeatable execution even for fuzzing inputs, while preserving access to advanced \ac{CPU} features like speculation, virtual memory, and \ucode engine.

\begin{figure*}
  \centering
  \includegraphics[width=1\textwidth]{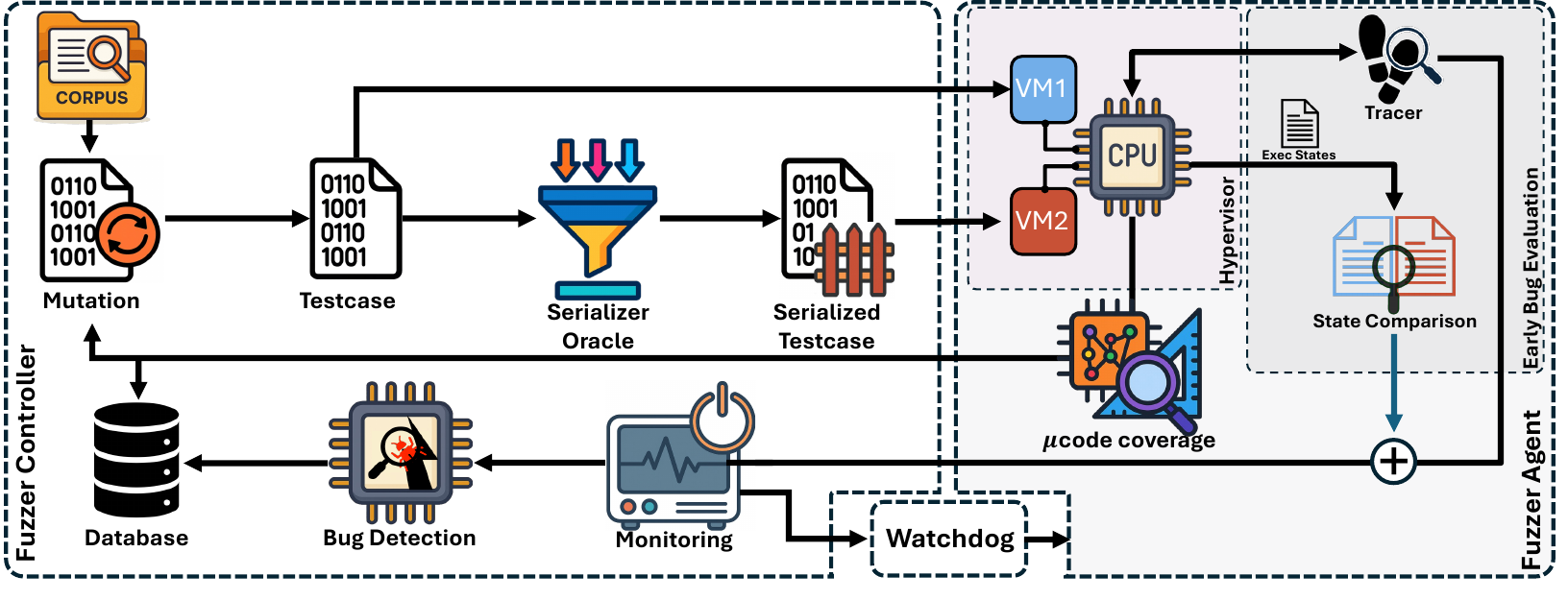}
  \caption{\label{fig:architecture} High-level overview of \ufuzz architecture}
\end{figure*}

\section{Design}
\label{sec:design}

In this section, we begin with a high-level overview of \ufuzz's design. We then introduce the \ucode coverage metric and explain how it exposes \uarch execution for effective exploration. Next, we describe our low-overhead bare-metal hypervisor that isolates execution while supporting efficient \ucode coverage collection. Finally, we present our \textit{Serialization Oracle} for vulnerability detection and discuss how we address its practical challenges. 

\subsection{High-level Overview}
The high-level overview of \ufuzz is shown in Figure~\ref{fig:architecture}, structured into three high-level components: \textit{Fuzzer Controller}, \textit{Fuzzer Agent}, and \textit{Watchdog}. The \textit{Fuzzer Controller} runs off-target \ac{CPU} and orchestrates the fuzzing campaign: it maintains the test corpus, mutates test cases, generates serialized variants for each mutated input, and dispatches both the original and serialized tests to the \textit{Fuzzer Agent}. 

The \textit{Fuzzer Agent} runs on the target \ac{CPU}. It pulls test cases from the \textit{Fuzzer controller}, applies \ucode instrumentation that records \ucode coverage into \ac{RAM}, provisions isolated execution using \ufuzz's bare-metal hypervisor, and executes each test both as-is and as its serialized variant in separate \acp{VM}. After each run, the \textit{Fuzzer Agent} collects the final architectural state (general-purpose and control registers) via the hypervisor interface and forwards it to the \textit{Early Bug Evaluation} stage. If the final architectural state of two executions diverges, a tracer replays the testcase instruction-by-instruction, snapshotting post-instruction architectural state to localize the root cause. The suspicious trace is then sent to the \textit{Fuzzer Controller} for triage and archival. In parallel with \textit{Early Bug Detection}, the \textit{Fuzzer Agent} exports post-execution \ucode coverage to the controller to guide mutations and records it for final reporting.

\ufuzz decouples the fuzzer from the \ac{DUT}. 
This separation ensures that unexpected target behavior (e.g., hangs, crashes) does not stall the campaign: all fuzzing state, collected coverage, and detected issues are durably stored. Upon detecting such failures, the \textit{Fuzzer Controller} triggers the \textit{Watchdog} that hard-resets the target \ac{CPU}, power-cycles the \textit{Fuzzer Agent}, and restores the target \ac{CPU} to a known good state.

\subsection{Microcode-Level Coverage Metric}
\label{sec:design:microcode}

\ufuzz introduces a novel feedback signal, \ucode coverage, which enables internal visibility into instruction execution at the \uops level. This feedback tracks which \ucode addresses are executed during test case execution, enabling the fuzzer to probe complex \uarch behaviors that are invisible to traditional architectural feedback.

As outlined in Section~\ref{cha:bg}, modern x86 processors execute complex instructions by translating them into sequences of \uops, stored in \ucode \ac{ROM}. To support post-silicon reconfiguration, \acp{CPU} also includes a writable patch \ac{RAM} and a redirection mechanism known as the \textit{hook table}. Upon instruction decode, the \textit{hook table} determines whether to use the default \ucode \ac{ROM} or redirect execution to a patch stored in \ac{RAM}. \ufuzz repurposes this reconfiguration mechanism to inject lightweight probes into the \ac{CPU}'s \ucode execution path. When an instruction executes, each associated \uop triggers a lightweight logging operation that increments a counter located in physical memory (\ac{RAM}). These counters are indexed by hook index, which can be later mapped to \ucode addresses, producing a precise execution profile that records which \uop is executed and the corresponding frequency. The \ucode coverage serves as a fine-grained feedback signal, enabling \ufuzz to prioritize inputs that explore new \uarch behaviors. To ensure that \ucode instrumentation does not affect \ac{CPU} functionality (Challenge 2 in Section \ref{sec:challenges}) \ufuzz preserves architectural state and control flow, and leverages unused or non-critical \ucode registers to avoid interfering with the instruction's intended behavior.

At runtime, the \textit{Fuzzer Agent} installs instrumentation by writing patched triads into \ucode \ac{RAM} and configuring the \textit{hook table} before execution; upon completion, it retrieves coverage from a memory-mapped region and reports it to the \textit{Fuzzer Controller}. By integrating with the processor's native \ucode patch infrastructure, \ufuzz provides \uarch introspection without emulation, turning the \ac{CPU} into a gray-box fuzzing target with \uarch visibility.

\subsection{Why Microcode Coverage?}
\label{sec:ucode-importance}
Architectural feedback, e.g., register differences and instruction coverage, observes only committed instruction outcomes and overlooks semantically distinct \uarch behaviors. A single x86 opcode, which is implemented by a \ucode routine and contains guarded control-flow (condition on \textit{VMX}/\textit{SMM} state, availability/health of internal units, retry/loop paths, fallbacks), can traverse different paths depending on \uarch state. For example, as illustrated by the \ucode pseudo code for \texttt{RDRAND} in Algorithm~\ref{alg:rdrand}. Lines 1, 2, and 5 gate behavior on these conditions: the same instruction may return a hardware random value, trap to \textit{SMM}, raise \#UD, or exit to a hypervisor while executing an identical instruction, yet architectural coverage reports “covered” after the first observation and provides no signal about which internal condition remains unexplored. In fuzzing terms, relying on \ac{ISA}-level outcomes is akin to function-coverage in software fuzzing~\cite{feedbackfuzzing}: it saturates early and fails to separate interesting edges inside the implementation.

We introduce the \ucode coverage to address this visibility gap (Challenge 1 in Section \ref{sec:challenges}) by exposing path-sensitive signals from the \ucode engine. By logging which \uop execute and with what multiplicity, it distinguishes inputs that exercise different internal paths for the same instruction. This yields a metric that guides the fuzzer to prioritize seeds that unlock new \ucode edges or change execution counts. 
In practice, \ucode coverage plays the role that edge coverage plays in software fuzzing~\cite{feedbackfuzzing}: it provides the fine-grained, state-sensitive feedback necessary to explore the CPU's \uarch functionality.

\begin{algorithm}
  \caption{Pseudocode summary of the \texttt{RDRAND} \ucode behavior.}
  \label{alg:rdrand}
  \begin{algorithmic}[1]
    \If{not in SMM}
      \If{LOCK prefix used}
        \State \textsc{Raise(\#UD)} 
      \EndIf
      \If{VMX}
        \State \textsc{Exit to SMM or VMX monitor}
      \EndIf
    \EndIf
    \State $number \leftarrow$ \textsc{HardwareRngGen}()
    \State $dst \leftarrow$ \textsc{ZeroExtend}($number$)
    \State \texttt{EFLAGS.CF} $\leftarrow$ \textsc{Set if $number \neq 0$}
  \end{algorithmic}
\end{algorithm}

\subsection{Hypervisor-Based Execution Isolation}
\label{sec:design:hypervisor}
To ensure deterministic execution and full isolation of fuzzing inputs (Challenge 4 in Section \ref{chal:nondeter}) \ufuzz introduces a lightweight, custom type-1 hypervisor~\cite{hypervisorstype} that encapsulates each test case within a dedicated virtual \ac{CPU} instance.
Unlike traditional virtualization systems designed for long-lived guest operating systems, \ufuzz's hypervisor is optimized for single-purpose, short-lived fuzzing executions with strict guarantees on state control and containment. For every test case, the fuzzer initializes a new virtual \ac{CPU} configured with a fixed architectural state. This includes the general-purpose registers, control and system registers, the instruction pointer, and segment selectors. The goal is to ensure that all inputs begin execution from an identical and defined \ac{CPU} state, eliminating residual effects from prior tests.

To enforce memory isolation, the hypervisor exposes a restricted, virtualized memory layout to the \ac{VM} using hardware-assisted paging. 
Control over memory permissions ensures that critical configuration structures remain immutable and that fuzzing inputs cannot overwrite or reuse memory in unintended ways. The hypervisor is responsible for intercepting instructions that may affect host system configuration, access IO ports, 
result in non-deterministic execution, and intercept non-maskable interrupts. 
To prevent non-termination, the hypervisor enforces a maximum execution time. Tests that exceed their budget are preempted with a forced \ac{VM} exit, ensuring that infinite loops cannot degrade the throughput or availability of the fuzzing campaign.
After the \ac{VM} terminates, the hypervisor extracts the architectural state of the virtual \ac{CPU} for bug detection. 
The hypervisor is tightly integrated into \ourname with low overhead, detailed in Section~\ref{sec:impl-hypervisor}.

\subsection{Differential Testing with Serialization Oracle}
\label{sec:design-serialization}

To detect bugs and vulnerabilities without a formal reference model, \ufuzz employs a differential strategy, inspired by the Zenbleed~\cite{zenbleed}, that compares each input program $P$ with a serialized variant $Q$, constructed to be semantically equivalent. The central idea is that if a processor is functionally correct, $P$ and $Q$ should produce the same architectural outcome. Any deviation implies a hidden inconsistency in \uarch behavior. The transformation that produces $Q$ enforces instruction-level serialization by inserting \textit{fence instructions} between every instruction to suppress speculation and reordering.

However, this transformation alters code layout, breaks relative addressing, and disrupts control and data flow. To overcome these challenges, \ufuzz incorporates systematic transformation strategies that preserve instruction semantics despite structural modification. 
For example, instruction pointer relative immediate memory accesses are adjusted by changing the opcode's operand to access the same memory location, like in the original program. 

Additionally, \ufuzz detects misaligned or non-standard control transfers, such as jumps into instruction bodies. In these cases, the program is unrolled into isolated fragments that can be serialized independently and safely joined (see Appendix~\ref{appendix:misaligned-jumps}).
This Serialization Oracle lets \ufuzz expose bugs rooted in \uarch features, such as speculation, transient state retention, hidden instruction, without a reference model (Challenge 2, Section~\ref{sec:challenges}).
The implementation of this Serialization Oracle is explained in Section~\ref{sec:impl-differential-testing}.

\section{Implementation}
\label{sec:implementation}

In this section, we detail the implementation of the \ufuzz framework. We first present the system architecture and deployment approach, followed by our \ucode instrumentation technique. We then describe our patching and coverage collection strategy, the hypervisor-based execution environment, our fuzzing input generation and mutation methods, and finally our differential testing approach with serialization oracle.

\subsection{System Architecture and Deployment}
\label{sec:impl-overview}

As outlined in Section~\ref{sec:design}, \ufuzz is comprised of three components: the \textit{Fuzzer Controller}, \textit{Fuzzer Agent}, and \textit{Watchdog}, implemented in Rust and x86-64 assembly. For target \ac{CPU}, we utilized a Gigabyte \textit{GB-BPCE-3350C}~\cite{board} Mini-PC containing an \textit{Intel N3350} (\textit{Goldmont} \uarch, \textit{CPUID} \texttt{0x506ca}).
The target \ac{CPU} does not natively expose interfaces for custom \ucode patches. We first red-unlock the processor following established procedures for this processor family~\cite{intelptxpoc,alaoui2019ptxe,ermolov2017hack,undocumented-instructions,CustomProcessingUnit}. Upon successful red-unlocking, undocumented instructions become available that enable \ac{CRBUS} access (\texttt{udbgrd} and \texttt{udbgwr}), which are essential for \ucode manipulation during fuzzing. This is a one-time manual procedure per target device.

Both the \fcontroller and \fwatchdog are hosted Rust applications running on Raspberry Pi 4 platforms with \textit{NixOS}~\cite{nixos}. The \fcontroller serves as the stateful core of the \ufuzz framework, generating test cases, performing mutations, communicating with the \fagent to deploy tests and collect \ucode coverage, and coordinating with the \fwatchdog to manage the target \ac{CPU}. The \fwatchdog functions as a remote keyboard and storage device for the \fagent, responsible for hard-resetting the target, booting it to \ac{UEFI}, and starting the \fagent executable. It exposes a \ac{REST} \ac{API} that allows the \fcontroller to issue control commands, for instance, upon detecting communication timeouts or unresponsive behavior, the \fcontroller triggers a reset command that causes the \fwatchdog to physically cycle power via probes soldered to the target motherboard. The \fagent deploys as a standalone \ac{UEFI} application on the target \ac{CPU} to eliminate \ac{OS} interference. It injects \ucode instrumentation and custom patches, executing fuzzing test inputs, collecting coverage, and communicating with the \fcontroller over \ac{UDP} for task coordination, test input delivery, and \ucode coverage collection.

This modular hardware layout ensures that the \fcontroller and \fwatchdog remain unaffected by crashes or lockups in the target \ac{CPU}, enabling long-term autonomous operation over days or weeks.

\subsection{\ucode Instrumentation and Patching}
\label{sec:impl-microcode}
\ufuzz enables runtime instrumentation of \uop by deploying patched \ucode sequences and configuring \ucode \textit{hook table}. As outlined in Section~\ref{sec:design}, the goal is to guide the fuzzer toward exploring all possible paths within \ucode implementations-effectively covering all reachable \ucode addresses. To achieve this, we instrument the target \ucode addresses by adding entries to the \ucode \textit{hook table} that redirect execution to coverage collection logic, then resume from the original address upon completion.

To preserve target \ac{CPU} functionality, coverage collection minimizes state changes. Our instrumentation first saves a required subset of the general-purpose registers in the \ac{CPU}'s staging buffer~\cite{CustomProcessingUnit}, records the coverage event and updates the in-memory coverage map, restores the saved state, executes the overwritten \uop, and finally jumps to the originally intended next \ucode address to continue normal execution. To the best of our knowledge, the \texttt{write-to-staging-buffer} is the only known \uop capable of storing register values into memory without needing an additional register operand. Since the staging buffer is shared among all \ucode routines, \ufuzz selects an address location normally used by the \texttt{udbgwr} instruction, ensuring a minimal state impact~\cite{CustomProcessingUnit,undocumented-instructions}. This approach ensures that the target's architectural and \uarch behavior remains unaffected.

Since the instrumentation only uses a subset of the available general-purpose registers, we only save and restore the state of these registers, keeping the instrumentation overhead profile low while maintaining correctness.

Our reverse engineering analysis uncovered that the \ucode engine implements address hook redirection through a paired addressing mechanism. Specifically, when a redirection hook is configured for an even address $S$ to destination $D$, the \ucode engine automatically redirects the corresponding odd address $S+1$ to destination $D+1$ without requiring an explicit hook entry.
Consequently, determining whether the hook was triggered by an even or odd \ucode address becomes critical for proper execution flow control and coverage calculation. To handle this differentiation, \ufuzz 
places an immediate jump at the hook redirect destination $D$, to ensure that \ucode execution originating from address $S$ will execute this instruction, while execution triggered from $S+1$ will skip it and continue to subsequent instructions. This mechanism enables \ufuzz to accurately differentiate and log the specific entry address ($S$ or $S+1$) that triggered the hook.

Since applying a \ucode hook redirection overwrites the processor's native \ucode at addresses $S$, $S+1$, \ufuzz manually inlines the corresponding original \ucode instructions at the conclusion of the injected sequence. Following this restoration, an immediate jump instruction is inserted to resume execution at the appropriate continuation address ($S+1$, $S+2$, or jump target if the original \uop was a jumping instruction).

For instrumenting a new \ucode address to enable coverage collection, both the inlined original \ucode instructions and the jump instruction must be updated accordingly. Given that each of the 16 \ucode hook registers maps to a pair of addresses (even and odd), complete reconfiguration necessitates writing 144 distinct values\footnote{$16\times(2\times4+1)$ Two triads and the \textit{hook table} entry for each \textit{hook table} entry.} through the \texttt{udbgwr} instruction.
To minimize the computational overhead associated with this setup process, \ufuzz implements a custom \ucode utility function that accepts a memory address parameter and performs the setup of all active \textit{hook table} entries within a single instruction decode cycle. Consequently, reconfiguring all hook targets requires only a single \texttt{udbgwr} invocation per test case iteration, substantially reducing the instrumentation overhead and improving overall fuzzing efficiency.

\subsection{Coverage Collection Strategy}
\label{sec:impl-cov-schedule}

Our target platform is constrained to only $16$ \ucode hooks, which limits the instrumentation coverage to $32$ addresses per instrumentation round. Given the approximately $32k$ (exact number \texttt{0x7C00}) \ucode address space, collecting comprehensive coverage for all \ucode addresses within a single fuzzing test case requires \ufuzz to execute the test case $\frac{0x7C00}{32}=992$ times. This baseline approach introduces significant performance overhead.

To address this fundamental limitation, \ufuzz implements the following optimized coverage collection scheduling mechanism that significantly reduces execution overhead while maintaining comprehensive coverage collection capabilities. During the initial fuzzing phase, \ufuzz instruments all x86 instruction entry points residing below the \texttt{0x1000} address range~\cite{undocumented-instructions}. 
However, these instruction entry addresses are aligned to 8 addresses, resulting in a total of $\frac{0\text{x}1000}{8}=512$ entry locations. Using $16$ hook registers, each fuzzing input is executed $\frac{512}{16}=32$ times to achieve initial coverage.

Subsequently, static analysis and \ucode disassembly techniques can be used to extract basic blocks from \ucode.
A basic block is defined as a set of \uops that once the first \uop of the block executes, unconditionally execute till the last \uop. Further, the only entry point is the first \uop, while the last \uop is always the last executed instruction of the basic block~\cite{bb}.

When instrumenting one \ucode address of any basic block, the coverage can be propagated to all addresses that are part of the block.
Furthermore, the set $\Phi(y)$ captures all conditionally reachable successors of a basic block $y$. Only these successors within $\Phi(y)$ are selected for subsequent instrumentation and test re-execution when the coverage of $y$ is non-empty. This selective approach eliminates unnecessary iterations while ensuring complete and fine-grained coverage collection.

Overall, this optimization strategy significantly reduces the computational overhead from $992$ times re-executions required in the baseline approach to at best $32$ iterations, plus the number of conditional branches encountered during \ucode execution, which benefits us with at best $31\times$ less overhead.

\subsection{Hypervisor-Based Program Execution}\label{sec:impl-hypervisor}

We implemented a type-1 hypervisor, running on top of \ac{UEFI}, providing fuzzing input isolation using the \textit{Intel VMX} virtualization technology. We provide a reproducible fuzzing execution harness by initializing a new virtual \ac{CPU} instance, memory isolation, and executing controls. First, we instantiate a fresh virtual \ac{CPU} by initializing the \ac{VMCS}, which is the main configuration structure for \textit{Intel VMX}. We set the initial values of stack and program counter registers, set up \ac{VM}-execution control setting~\cite{intel-system-manual}, and set the memory translation-related registers. Using \textit{Extended Page Tables}~\cite{ept}, we provide the same virtualized view of main memory to each fuzzing input, split into three parts: execute-only, read-write, read-only, initialized once. Before invoking the fuzzing input (in an execute-only region), the read-write region is zeroed to purge transient state from prior executions. The read-only region is only initialized once at fuzzing start, containing required memory structures for x86 instruction execution, like a \ac{GDT}, and a \ac{TSS} data structure. Using the \ac{VM}-execution control settings, we configure the fuzzing guest \ac{VM} to exit on problematic actions like executing an instruction that introduces nondeterminism, interaction with hardware components, arrival of an external interrupt, and the \ac{VM}'s timeout budget. The hypervisor may handle each event and decide to stop or continue fuzzing input execution. Using the timeout, we prevent loops in the fuzzing input from stalling the fuzzing process. When finishing executing a fuzzing sample, the architectural state of the virtual \ac{CPU} instance is captured and sent to the fuzzing controller into the bug-detection pipeline.

\subsection{Fuzzing Input Generation and Mutation}
Each fuzzing campaign starts with an initial corpus, comprising randomly generated byte sequences or valid instruction gadgets extracted from software libraries, such as \textit{libcxx}. Valid instructions are extracted from software libraries and randomly concatenated until a configurable size threshold is achieved.

Input mutation is performed using either a custom genetic algorithm guided by coverage and execution-depth heuristics, or established mutators (e.g., Havoc) from the \textit{libafl} library~\cite{libafl}. Our custom mutation engine implements random 1-to-8 byte mutations and cross-over operations on the fittest fuzzing samples. The fitness evaluation of a given sample is computed using the \ucode coverage metric, combined with the ratio of executed bytes to total generated bytes. We preserve the top-$k$ samples per generation and utilize them to seed subsequent generations.
Consequently, the generated fuzzing inputs may comprise valid \ac{ISA} instruction sequences or arbitrary invalid byte sequences, all evolved through \ucode coverage feedback mechanisms

\subsection{Differential Testing with Serialization Oracle} \label{sec:impl-differential-testing}

To implement the serialization oracle discussed in Section~\ref{sec:design-serialization}, we encounter a non-trivial challenge: program layout changes cause relative addressing instructions to reference incorrect locations. We address this by analyzing each instruction's behavior according to the modified program layout. Our oracle maintains a list of all relative addressing instructions, extracts these from the original test case, and recomputes their offsets for the transformed program, following established binary rewriting methodologies~\cite{duck2020binary}. However, fuzzing inputs are comprised of randomly generated byte sequences rather than compiler-generated code, enabling jump or control flow instructions to target mid-instruction locations (Appendix~\ref{appendix:misaligned-jumps}). Introducing serialization in such cases creates non-equivalent execution between serialized and non-serialized variants, generating numerous false positives during vulnerability detection.
We solve this problem by adapting Superset Disassembly~\cite{binaryRewriting} techniques, unrolling the original program to discover additional instruction gadgets. However, we utilize a different relocation approach: rather than relocating entire blocks, we relocate individual instructions to new locations, inserting fence instructions between each relocated instruction to ensure deterministic execution ordering.

A corner case involves self-modifying code and fuzzing test cases that attempt to alter their own instructions during execution. Since self-modifying code dynamically changes program execution flow at runtime, it can introduce new relative addressing instructions that static binary rewriting cannot anticipate or handle, as rewriting occurs before program execution.
We address this limitation by restricting self-modifying behavior in fuzzing test cases through execute-only page permissions. The hypervisor detects and interrupts any fuzzing input that attempts to write instructions to execute-only memory pages, preventing dynamic code modification that would compromise our serialization oracle's correctness.

\section{Evaluation}
\label{sec:evaluation}

In this section, we first discuss \ufuzz's automatic detection of $\mu$Spectre~\cite{uSpectre} class vulnerabilities. We then present a specialized use case where \ufuzz targets speculative execution leakage at the \ucode level, leading to two newly discovered vulnerabilities and three new and interesting findings related to speculative behavior in \ucode.
Next, we evaluate \ucode coverage effectiveness by comparing total achieved coverage, coverage acquisition speed, and the impact of random versus carefully crafted seeds and mutation engines. Since existing works~\cite{silifuzz} lack clear coverage definitions, direct comparison is not feasible; instead, we compare against a baseline x86 fuzzer without coverage guidance to demonstrate \ufuzz's advantages over the class of existing approaches. Each coverage experiment runs for $48$ hours with at least three repetitions to minimize noise. Finally, we analyze \ufuzz's performance characteristics and \ucode coverage overhead.

\subsection{F1. $\mu$Spectre Vulnerabilities Detection}
\label{sec:eval-general-findings}
The $\mu$Spectre vulnerability class, discovered by Mosier et al.~\cite{uSpectre}, exploits \ucode-level speculative execution where branches are statically predicted (taken, not-taken, or stalled). This design choice enables data leakage through mispredicted branches that continue speculative execution across \xx\ instruction boundaries, allowing subsequent instructions to leak information. During our fuzzing campaign, \ufuzz automatically detected this vulnerability class during \ucode instrumentation (\textbf{F1}). The \ucode instrumentation itself does not modify the test case's instruction sequence. However, the act of instrumenting \ucode can trigger speculative execution behaviors at \ucode-level that wouldn't occur naturally, for instance, by having \ucode-level branches that are statically predicted. When the same test case is re-executed with serialization fences (such as \texttt{LFENCE} instructions), these fences prevent speculation from reaching the following instruction. The divergence in architectural state between the speculative and non-speculative executions reveals the speculative behavior, allowing \ufuzz to automatically identify instances of the $\mu$Spectre vulnerability class without prior knowledge.

\subsection{Use Case: Fuzzing x86 Microcode for Speculative Leakage}
\label{sec:eval-specleak}

Speculative execution enables processors to transiently execute instructions (\uop) before preceding control flow is resolved. While effective for performance, this behavior creates the potential for sensitive information to leak through microarchitectural side effects~\cite{kocher2020spectre,zenbleed,Moghimi2023,lipp2018meltdown}. Prior work has largely focused on instruction-level speculation behaviors~\cite{rostami2024lost,hur2022specdoctor,zenbleed}. \ufuzz opens a new frontier: fuzzing the \ucode sequences that execute transiently within speculative windows to discover information leakage or side channels originating at the \ucode level. Rather than modifying the branch predictor behavior directly, we built a reusable \ucode template (Listing~\ref{lst:spec_template}) that creates a speculative execution context. We leverage an existing \uop, \texttt{UJMPCC\_DIRECT\_NOTTAKEN\_CONDNZ} (Opcode: \texttt{0x151}) which is designed to always predict ''not taken'' and initiate speculative execution at the \ucode level~\cite{uSpectre}. We use this created speculative path as a stable speculative entry point for \ucode fuzzing.

\noindent \textbf{Speculative Template Construction:} We reserve a region of the \ucode patch RAM as the speculative body (see Listing~\ref{lst:spec_template}). During each fuzzing iteration, \ufuzz injects a randomized or guided sequence of \uops into this region. These \uops execute speculatively because, as discussed, \texttt{UJMPCC\_DIRECT\_NOTTAKEN\_CONDNZ} always will be mispredicted and create a speculative window. These \uops cause \uarch effects that should be rolled back by design upon the end of the speculative window.
\noindent
\textbf{Leakage Observation:} Although the speculative path does not commit architecturally, its execution may leave detectable architectural and \uarch traces. \ufuzz detects such leakage by comparing the architectural state before and after execution of the template~(see Listing~\ref{lst:spec_template}). This enables the detection of persistent side effects introduced solely by speculative \uops execution.

This use case demonstrates how the \ufuzz framework enables targeted fuzzing of speculative behavior at the \ucode level, uncovering security-relevant leakages that are invisible to architectural-level fuzzers. Below, we summarize key findings made possible through this capability.

\begin{lstlisting}[caption={Speculative \ucode window using forced misprediction template}, label={lst:spec_template}, numbers=none]
<entry>
tmp2 := ZEROEXT_DSZ64(0xabab)
tmp0 := ZEROEXT_DSZ64(0x1000)
tmp1 := LDPPHYS_DSZ32_ASZ16_SC1(tmp0)
tmp0 := SUB_DSZ64(tmp0, tmp1)

; Speculative branch | misprediction forced
UJMPCC_DIRECT_NOTTAKEN_CONDNZ(tmp0, <taken>)
; Speculative Window Start
; --- INSERT MICRO-OPERATION HERE ---
rax := ZEROEXT_DSZ64(0xdead)
NOPB
NOP SEQW SYNCFULL
NOPB
; Speculative Window Ends

<taken>
unk_256() !m1 SEQW LFNCEWAIT, UEND0
\end{lstlisting}

\noindent
\textbf{F2. Speculative writes to the \ac{CRBUS} persist after the speculative window is discarded.} 

We discovered that certain \uops writing to the \ac{CRBUS} within speculative execution windows persist their side effects even when the \ac{CPU} subsequently discards the speculated execution. This leads to incorrect or unintended state changes that violate speculative execution's rollback guarantees.
For example, executing micro-op \texttt{0x4292180220} \texttt{MOVETOCREG\_DSZ64(rax, 0x692)} writes to CRBUS address \texttt{0x692}, which controls the \ucode \textit{hook table} activation. When \texttt{rax} contains \texttt{0x1} (deactivate), all \ucode patched, including security patches, are disabled, effectively bypassing current \ucode security updates, even when executed speculatively and subsequently rolled back.
We provide a comprehensive list (See Appendix~\ref{apdix:microcode_hang}) of \ucode operations whose effects persist during speculative windows despite execution rollback by triggering unrecoverable \ac{CPU} lockups that constitute denial-of-service conditions. We categorize their behaviors as either reliable (StableTimeout), always causing a lock-up or unreliable (Unstable), sometimes causing a denial-of-service.

\noindent
\textbf{F3. Speculative updates to the segment selector caches persist after the speculative window is discarded.} 
Our analysis reveals a critical vulnerability
in segment selector cache management during \ucode-level speculative execution. When a \uop targeting segment selector caches (opcode \texttt{0xc6b} - \texttt{WRSEGFLD}) executes within a speculative window, their state modifications persist even when the triggering branch is subsequently mispredicted and the speculative execution should be discarded.
This vulnerability can be exploited when default \ucode implementations or security updates contain segment selector cache writes that execute speculatively but fail to undergo proper rollback following branch mispredictions. Critically, these persistent writes can bypass permission checks that would normally guard \uop execution. An adversary capable of executing code on the target processor can exploit this behavior across all privilege levels, potentially achieving denial-of-service conditions, privilege escalation to ring-0, or unauthorized memory access.
We demonstrate this vulnerability using \uop \texttt{0xc6b26000037} \texttt{WRSEGFLD(tmp7, GDT, BASE)} within our speculative execution template (Listing~\ref{lst:spec_template}). This operation writes the \texttt{tmp7} register value into the segment selector cache as the \ac{GDT} base address. The unauthorized modification can be detected through two mechanisms: indirectly via \ac{CPU} crashes when the target memory contains invalid \ac{GDT} structures, or directly through the Intel \textit{VMX} hypervisor \ac{API} that permits segment cache inspection and manipulation in virtualized environments.

\noindent
\textbf{F4. Microcode-implemented instructions terminate speculation.} Our analysis shows that speculative execution at the instruction level terminates when encountering instructions requiring \ucode implementation and the instruction cache is empty. Specifically, when a speculative window initiates, for example, following a mispredicted call-return sequence, the first instruction dispatched to a \ucode sequencer halts further speculative execution. This behavior indicates that \ucode dispatch boundaries function as implicit speculation barriers within \textit{Intel}'s processor implementation.
This architectural characteristic creates a timing-based observable that adversaries can exploit for information disclosure. An attacker can leverage these timing variations to infer sensitive information about program execution paths.

\noindent
\textbf{F5. Speculative $\mu$ops leave traces on performance counters.}
Our analysis shows that certain \uops executed speculatively produce measurable side effects on architecturally visible performance counters. For instance, speculative execution of the undocumented \texttt{UNK\_256} \uop increments the \ucode sequencer counter (\texttt{MS\_DECODE.MS\_ENTRY}) even when executed within speculative windows that are subsequently discarded. This behavior indicates that non-retired \ucode execution paths can leak internal processor state through architecturally observable side-channel artifacts. \textit{Intel}'s \ac{ISA} documentation reveals that this behavior aligns with documented processor specifications, although it can leak information.

Overall, these findings demonstrate the effectiveness of \ufuzz for exploring speculative behavior at the \ucode-level, and its capability to uncover information leakage channels and safety violations not observable through conventional fuzzing techniques.

\subsection{Coverage Analysis}
In this section, we present the results on the effectiveness of \ucode coverage feedback for exploring the \ac{CPU} architecture. During this analysis, we fixed the mutation engine on the AFL Havoc engine due to its superior performance in all experiments compared to our custom mutation. Also, we evaluated the impact of fuzzing corpus selection by conducting experiments using two distinct corpus types: (1) \textbf{Random Corpus}, consisting of concatenated random bytes up to a fixed size limit, and (2) \textbf{Valid Corpus}, constructed by extracting unique \xx\ instructions from three widely-used software libraries: \texttt{libc++}, \texttt{libz}, and \texttt{libzip}.

\begin{figure}
  \begin{center}
  {\includegraphics[width=\columnwidth]{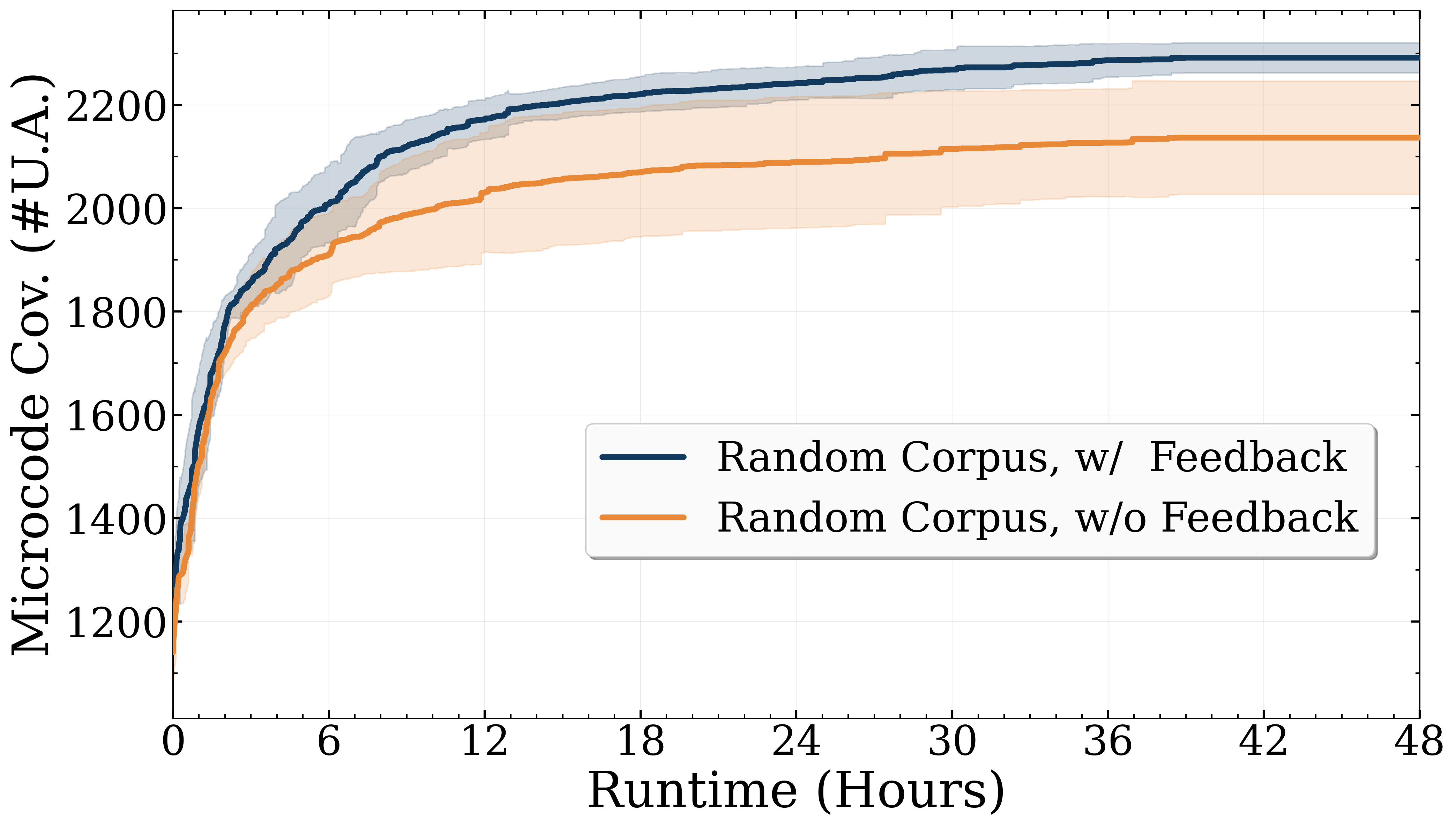}}
  \end{center}
\caption{\label{fig:afl_nc}Effectiveness of \ucode-coverage feedback on exploration with Havoc mutators and random corpus over time. \#U.A. stands for number of unique \ucode addresses.}
\end{figure}

\begin{figure}
  \begin{center}
  {\includegraphics[width=\columnwidth]{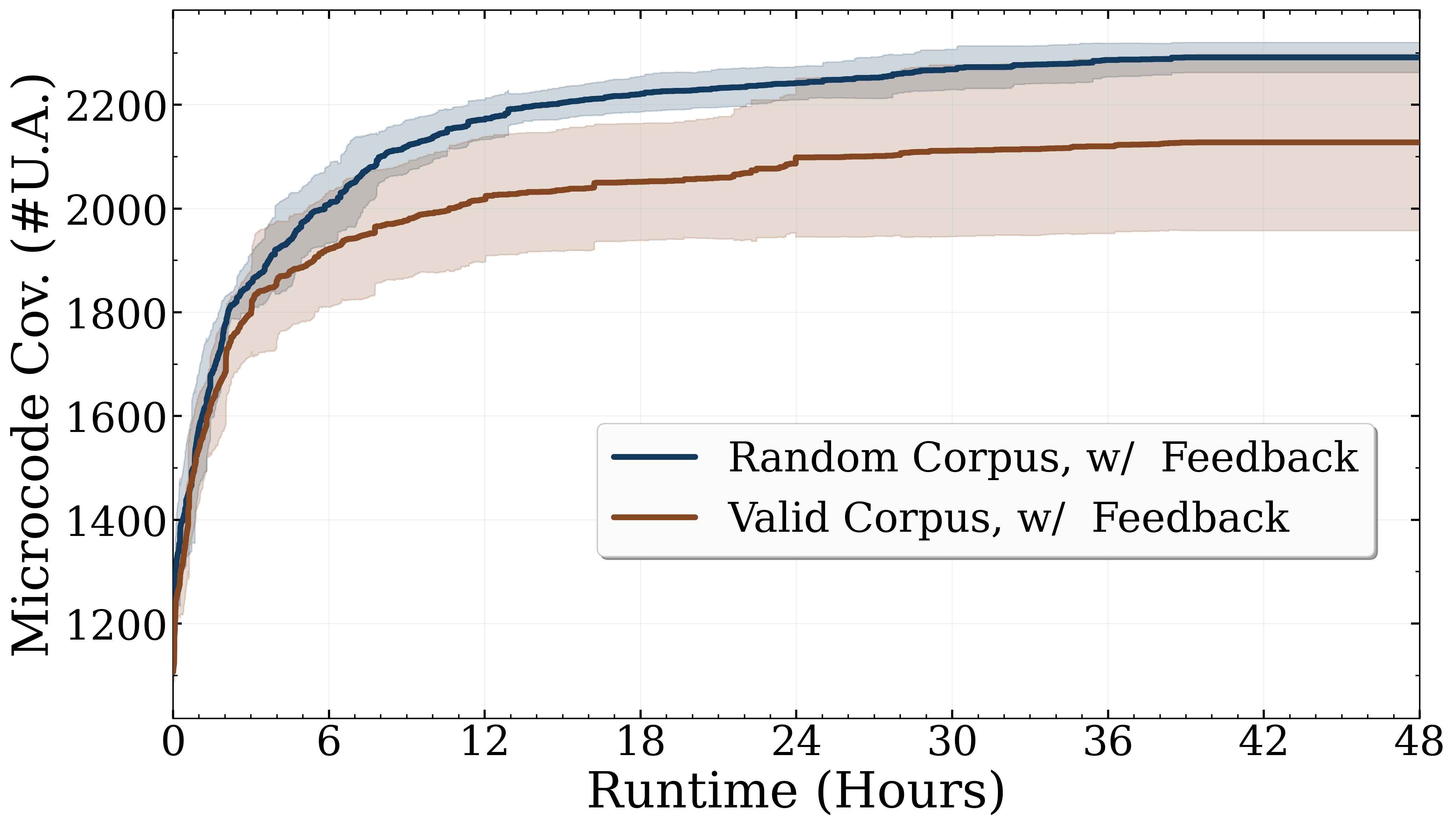}}
  \end{center}
\caption{\label{fig:afl_wf}Effectiveness of corpus on exploration with Havoc mutator and \ucode-coverage feedback over time. (\#U.A. stands for number of unique addresses.)}
\end{figure}

Figure~\ref{fig:afl_nc} shows the \ucode coverage metric over time when using random corpus in both fuzzing setups, with \ucode coverage feedback enabled in one configuration and disabled in the other. The fuzzing setup without feedback (\texttt{Random Corpus, w/o feedback}) represents the baseline \xx\ fuzzers that lack coverage metrics
, while the setup with feedback (\texttt{Random Corpus, w/ Feedback}) represents \ourname. As illustrated in the graph, \ourname outperforms baseline \xx\ fuzzing methodologies not only in final coverage achievement (across all runs) but also achieves the same coverage metric approximately $8\times$ faster on average. This demonstrates the effectiveness of \ucode coverage feedback in guiding the fuzzer toward exploring the \ac{CPU} design. To further evaluate the impact of corpus quality, we conducted experiments using both random corpus and valid corpus while maintaining feedback activation in both scenarios. As shown in Figure~\ref{fig:afl_wf}, average coverage achieved by \textit{Random Corpus} consistently outperforms \textit{Valid Corpus}, which our analysis attributes to the inherent diversity of random corpora generated from random bytes, whereas the valid corpora contain structured instructions that may exhibit redundancy across different corpus samples.

\begin{figure}
    \begin{center}
    {\includegraphics[width=\columnwidth]{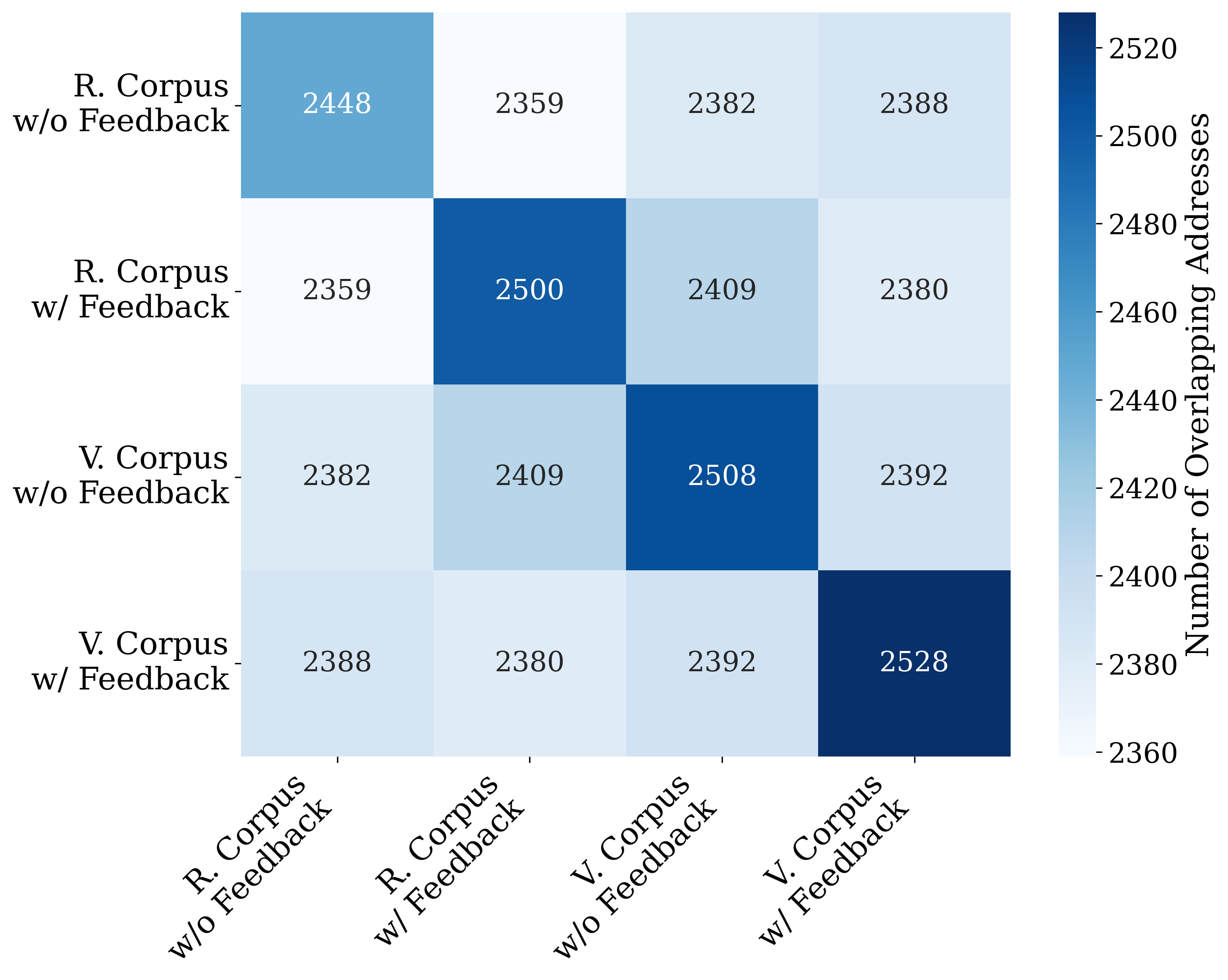}}
    \end{center}
    \caption{\label{fig:overlap_matrix}\ucode coverage overlap matrix. Addresses in row config overlapping with column config. \texttt{\#U.A.} stands for number of unique addresses. \texttt{R.} stands for random, and \texttt{V.} stands for valid.}
\end{figure}

\begin{figure}
    \begin{center}
    {\includegraphics[width=\columnwidth]{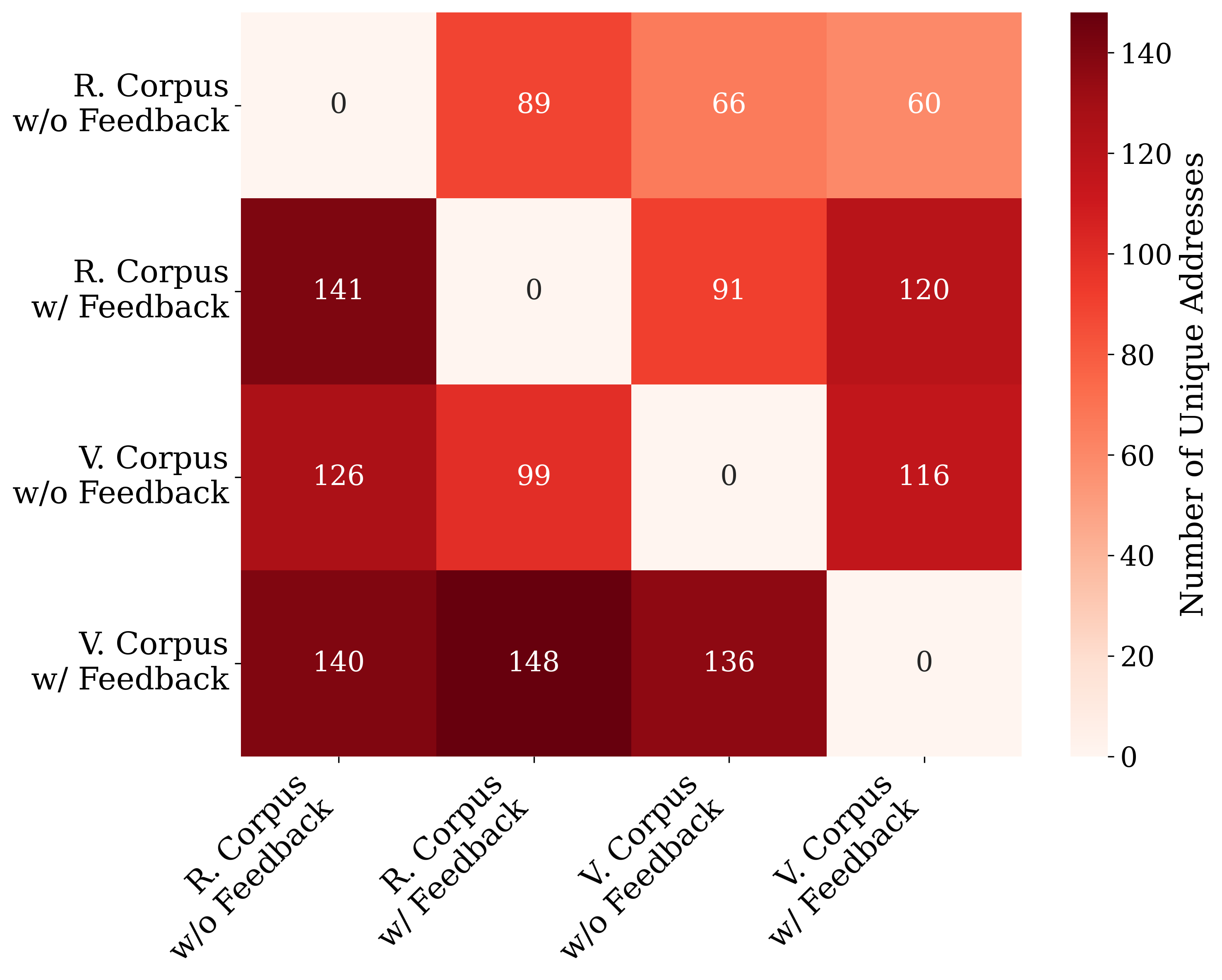}}
    \end{center}
    \caption{\label{fig:uniqe_matrix}\ucode coverage uniqueness matrix. Addresses in row config NOT in column config. \texttt{\#U.A.} stands for number of unique addresses. \texttt{R.} stands for random, and \texttt{V.} stands for valid.}
\end{figure}

\begin{figure}
  \begin{center}
  {\includegraphics[width=\columnwidth]{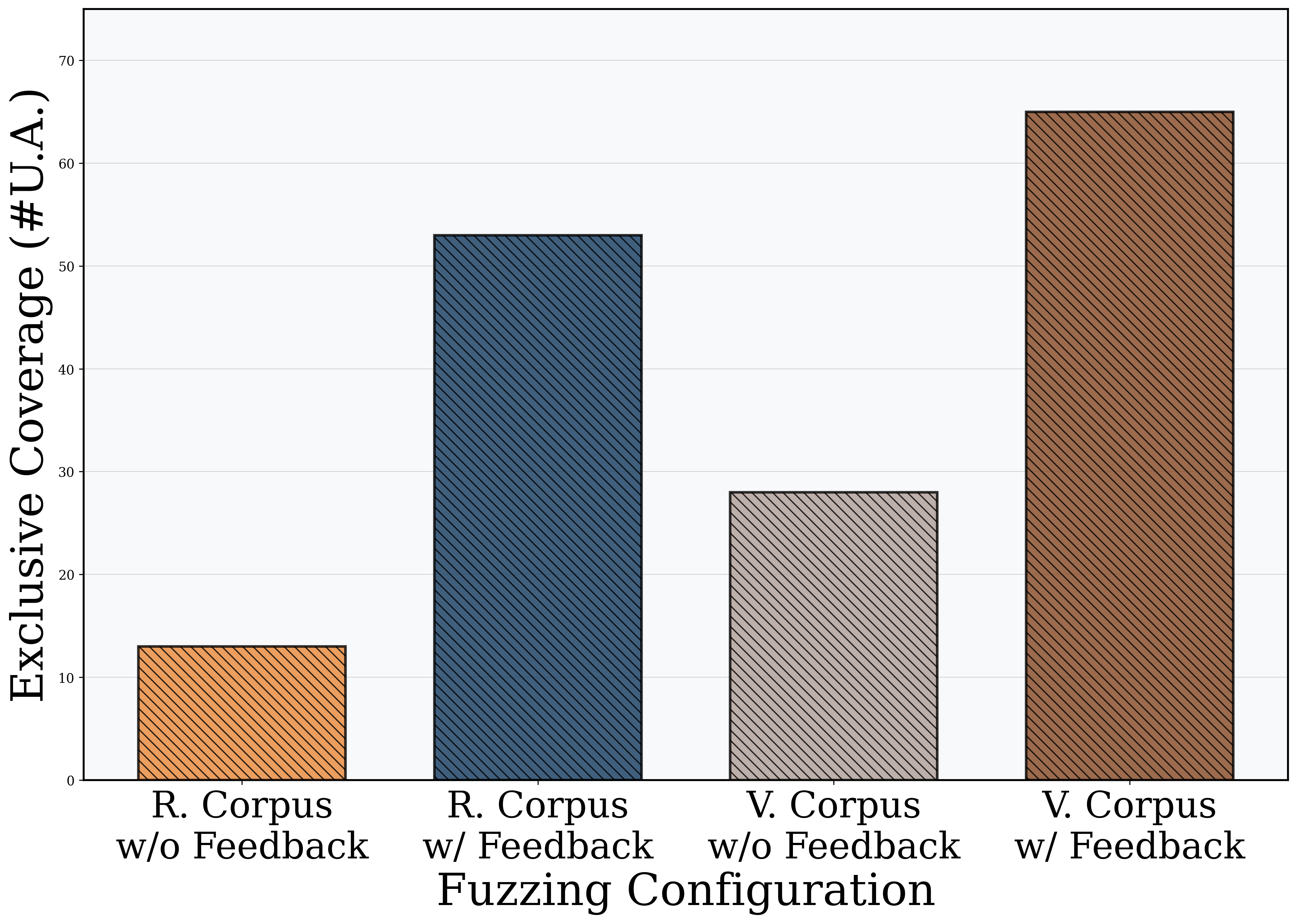}}
  \end{center}
\caption{\label{fig:exclusive} Exclusive \ucode Coverage for Different Fuzzing Configurations. (\texttt{\#U.A.} Stands for Number of Unique Addresses. \texttt{R.} Stands for Random, and \texttt{V.} stands for Valid)}
\end{figure}

However, Figures~\ref{fig:afl_nc} and~\ref{fig:afl_wf} only present the average number of unique \ucode addresses achieved in each fuzzing setup, lacking information about the total overlap between these addresses and the comparative effectiveness when combining all runs. To evaluate this metric comprehensively, we calculated the \ucode-coverage overlap matrix (Figure~\ref{fig:overlap_matrix}), \ucode-coverage uniqueness matrix (Figure~\ref{fig:uniqe_matrix}), and exclusive \ucode-coverage analysis (Figure~\ref{fig:exclusive}) across all runs for different setups.

Figure~\ref{fig:overlap_matrix} presents the \ucode-coverage overlap matrix, where diagonal cells indicate the total number of unique \ucode addresses achieved across all runs for each fuzzing setup, while off-diagonal cells represent the overlap between row and column configurations. As evident in Figure~\ref{fig:overlap_matrix}, \textit{Valid Corpus with Feedback} achieved $2,528$ unique addresses in total, surpassing all other configurations. Furthermore, configurations with \ucode-coverage feedback (\texttt{Valid Corpus, w/ Feedback} and \texttt{Random Corpus, w/ Feedback}) consistently outperform their respective counterparts without feedback (\texttt{Valid Corpus, w/o Feedback} and \texttt{Random Corpus, w/o Feedback}), demonstrating the effectiveness of our introduced \ucode-coverage feedback in exploring more unique \ucode addresses. Additionally, this analysis reveals that \textit{Valid Corpus} configurations achieve higher total unique address coverage across multiple fuzzing rounds (accounting for fuzzing reset effects~\cite{schiller2023novelty}) compared to \textit{Random Corpus} configurations. This finding reconciles the apparent contradiction in Figure~\ref{fig:afl_wf}, where \textit{Random Corpus} shows better average performance per run, but \textit{Valid Corpus} configurations achieve higher total coverage for all fuzzing runs.


Figure~\ref{fig:uniqe_matrix} shows the \ucode-coverage uniqueness matrix, where each matrix cell represents the number of unique \ucode addresses covered by the row fuzzing setup but not by the column setup. This visualization supports the same conclusions as the overlap matrix: fuzzing setups with \ucode-coverage feedback consistently achieve more unique \ucode addresses compared to their counterparts without feedback. Moreover, fuzzing setups with \textit{Valid Corpus} consistently achieve more unique \ucode addresses than \textit{Random Corpus} configurations, supporting our previous conclusions.

Finally, figure~\ref{fig:exclusive} illustrates the number of exclusive coverage points achieved by each fuzzing configuration compared to all other setups, specifically, how many unique \ucode addresses a fuzzing setup discovered that other setups failed to achieve. As demonstrated, configurations with \ucode-coverage feedback enabled consistently achieve at least $2\times$ higher exclusive \ucode-coverage points compared to any setup without \ucode-coverage feedback. This provides strong evidence for the effectiveness of \ucode-coverage feedback and \ourname in exploring the \ac{CPU} \uarch compared to existing baseline methodologies.

In total, across all our fuzzing campaigns, \ourname achieved coverage of \totalcovpoint unique \ucode addresses, representing \totalcovp of all hookable \ucode addresses (\possiblecovpoint) in the target CPU architecture.

\subsection{Performance Analysis}

\ourname introduces two primary sources of overhead: 1) Hypervisor-based execution environment (reset architectural state, memory, start end stop \ac{VM} execution), and 2) \ucode-coverage collection and fuzzing framework overhead like capturing the architectural state.
To quantify these, we instrumented \ourname to collect timing measurements from each source of overhead (see Table~\ref{tab:iteration-overhead}).
In comparison to the baseline approach~\cite{CustomProcessingUnit}, \ourname optimization significantly reduces instrumentation overhead. The baseline approach, when fully parallelized across 16 hooks, incurs an overhead of $566.490$ milliseconds to complete one fuzzing round, whereas \ourname completes the same task in only $18.274$ milliseconds, achieving an approximately $31\times$ overhead reduction in the best-case scenario.

\begin{figure}
\begin{center}
\resizebox{\columnwidth}{!}{%
\begin{tabular}{l|c|r|r}
\textbf{Fuzzing Step} & \textbf{Average Time} & \textbf{std dev} \\ \hline \hline
Hypervisor Setup arch-state & 164.173us & 3.317us \\ \hline
Hypervisor Setup memory & 149.833us & 3.943us \\ \hline
State Capturing & 18.404us & 1.600us \\ \hline
Coverage Setup & 183.054us & 86.030us \\ \hline
Coverage Collection & 55.594us & 2.808us \\ \hline \hline
\textit{Total Overhead (Single Execution Round)} & \textit{571.058us }& - \\ \hline \hline
\rowcolor{gray!20}
{Baseline total overhead} & 566.490ms & - \\ \hline
\rowcolor{gray!20}
\textbf{\ufuzz\ total overhead} & 18.274ms & - \\ \hline

\end{tabular}%
}
\caption{Timing Breakdown of Overheads in \ufuzz. \textit{std dev} Stands for Standard Deviation.}
\label{tab:iteration-overhead}
\end{center}
\end{figure}

\section{Discussion}\label{sec:disc}

\noindent\textbf{Portability to other CPU families.}
While the \ufuzz{} framework was evaluated on an \textit{Intel N3350 processor} (Goldmont \uarch, \texttt{CPUID: 0x506ca}), its applicability extends beyond this specific \ac{CPU}. The framework can be adapted to other Red-unlocked \textit{Intel} processors with minimal modifications, primarily adjusting \uarch-specific parameters. No modification is required for processors sharing the same microarchitecture or \ac{CPU} family.
Extending \ufuzz{} to other \ac{CPU} vendors or non-Red-unlocked \acp{CPU} is feasible if \ucode update interfaces remain accessible. When \ucode update capabilities are available, only the instrumentation logic requires adjustment to accommodate the target \ac{CPU}'s specific update procedures.
Recently, \textit{Google} researchers disclosed \textit{AMD}'s EntryBleed vulnerability~\cite{amd-sb-7033}, which enables loading malicious \ucode patches on \textit{AMD} processors. The instrumentation logic of \ufuzz{} could potentially be extended to exploit this vulnerability for patching \ucode updates in \textit{AMD} systems. However, unlike \textit{Intel}'s extensively reverse-engineered \ucode architecture, \textit{AMD}'s \ucode is still poorly understood and undocumented. Consequently, we defer \textit{AMD} \ac{CPU} fuzzing to future research efforts.

\begin{figure*}
\begin{center}
\resizebox{\textwidth}{!}{%
\begin{tabular}{l|l|l|l|l|l|l|l|l|l}
Year & Method                                 & Type                    & Target         & \begin{tabular}{@{}l@{}}Input\\generation\end{tabular} & ISA-Simulator  & Vulnerability detection           & Platform              & \begin{tabular}{@{}l@{}}
    $\mu$-arch\\feedback
\end{tabular} & \begin{tabular}{@{}l@{}}
    Fuzz-input\\restriction
\end{tabular}     \\ \hline \hline
2018 & RFUZZ~\cite{rfuzz}                     & pre-silicon             & RISC-V+        & Stochastic       & not-applicable & Assertion checking (i.d.)        & FPGA                  &—&—         \\ \hline
2021 & DIFUZZRTL\cite{difuzzrtl}              & pre-silicon             & RISC-V         & Stochastic       & yes            & Golden reference model (o.d.)     & FPGA                  &—&—         \\ \hline
2021 & EPEX~\cite{epex}                       & pre-silicon             & RISC-V         & Stochastic       & yes            & Equivalent program (i.d.)         & FPGA                  &—&—         \\ \hline
2022 & TheHuzz~\cite{thehuzz}                 & pre-silicon             & RISC-V+        & Stochastic       & yes            & Golden reference model (o.d.)     & Emulation             &—&—         \\ \hline
2022 & Cross-Level [...]~\cite{Bruns2022}     & pre-silicon             & RISC-V         & Stochastic       & yes            & Golden reference model (o.d.)     & Emulation             &—&—         \\ \hline
2023 & HyPFuzz~\cite{hypfuzz}                 & pre-silicon             & RISC-V         & Formal-assisted  & yes            & Golden reference model (o.d.)     & Emulation             &—&—         \\ \hline
2023 & PSOFuzz~\cite{psofuzz}                 & pre-silicon             & RISC-V         & Stochastic       & yes            & Golden reference model (o.d.)     & Emulation             &—&—         \\ \hline
2023 & MABFuzz~\cite{mabfuzz}                 & pre-silicon             & RISC-V         & Stochastic       & yes            & Golden reference model (o.d.)     & Emulation             &—&—         \\ \hline
2023 & MorFuzz~\cite{morfuzz}                 & pre-silicon             & RISC-V         & Template         & yes            & Golden reference model (o.d.)     & Emulation             &—&—         \\ \hline
2023 & SoCFuzzer~\cite{socfuzz}               & pre-silicon             & RISC-V         & Stochastic       & no             & Assertion checking (i.d.)         & FPGA+OS               &—&—         \\ \hline
2023 & ProcessorFuzz\cite{processorFuzz}      & pre-silicon             & RISC-V         & Stochastic       & yes            & Golden reference model (o.d.)     & Emulation             &—&—         \\ \hline
2023 & SurgeFuzz~\cite{surgefuzz}             & pre-silicon             & RISC-V         & Stochastic       & no             & Assertion checking (i.d.)         & Emulation             &—&—         \\ \hline
2023 & StressTest~\cite{stresstest}           & pre-silicon             & unknown        & Template         & yes            & Golden reference model (o.d.)     & Emulation             &—&—         \\ \hline
2024 & ChatFuzz~\cite{chatfuzz}               & pre-silicon             & RISC-V         & LLM-assisted     & yes            & Golden reference model (o.d.)     & Emulation             &—&—         \\ \hline
2024 & Cascade~\cite{cascade}                 & pre-silicon             & RISC-V         & BasicBlock         & yes            & Halting problem (i.d.)            & Emulation             &—&—         \\ \hline
2024 & FuzzWiz~\cite{FuzzWiz}                 & pre-silicon             & not-applicable & Stochastic       & not-applicable & Assertion checking (i.d.)        & Emulation             &—&—         \\ \hline \hline
\rowcolor{gray!20}
2021 & Osiris~\cite{osiris}                   & \textbf{post-silicon}   & \textbf{x86}   & Stochastic       & \textbf{no}    & \cellcolor{gray!40}Time measurement (o.d.)            & \cellcolor{gray!40}OS                    & \cellcolor{gray!40}no           & \cellcolor{gray!40}yes    \\ \hline
\rowcolor{gray!20}
2021 & SiliFuzz~\cite{silifuzz}               & \textbf{post-silicon}   & \textbf{x86}   & Stochastic       & yes            & \cellcolor{gray!40}Inter-device (o.d.)               & \cellcolor{gray!40}OS                    & \cellcolor{gray!40}no           & \cellcolor{gray!40}yes    \\ \hline
\rowcolor{gray!20}
2024 & RISCVuzz~\cite{RISCVuzz}               & \textbf{post-silicon}   & RISC-V         & Stochastic       & \textbf{no}    & \cellcolor{gray!40}Inter-device (o.d.)               & \cellcolor{gray!40}OS                    & \cellcolor{gray!40}no           & \cellcolor{gray!40}yes    \\ \hline
\rowcolor{gray!20}
2025 & \ufuzz                                 & \textbf{post-silicon}   & \textbf{x86}   & Stochastic       & \textbf{no}    & \cellcolor{gray!40}\textbf{Serialized oracle (i.d.)} & \cellcolor{gray!40}\textbf{Bare-metal}   & \cellcolor{gray!40}\textbf{yes} & \cellcolor{gray!40}\textbf{no}  \\

\end{tabular}%
}
\label{tab:related-work-all}
\caption{Comparison of \ac{CPU} fuzzers. \ufuzz\ is the first post-silicon \xx\ fuzzer that does not require an operating system (hence has no limitations regarding input generation) and uses a \textit{Serialized-oracle} model for bug detection. Further, it is the first general \xx\ fuzzer that does not require an \ac{ISA}-Simulation. The o.d. stands for output-driven. The i.d. stands for input-driven.}
\end{center}
\end{figure*}

\section{Related work}
\label{sec:related-work}

We have structured the result of our literature research in Table \ref{tab:related-work-all}. We have identified 19~hardware fuzzers~\cite{rfuzz,difuzzrtl,epex,thehuzz,Bruns2022,hypfuzz,psofuzz,mabfuzz,morfuzz,socfuzz,processorFuzz,surgefuzz,stresstest,chatfuzz,cascade,FuzzWiz,osiris,silifuzz,RISCVuzz} published in the last years that we categorized using the categories: \textbf{Type} (pre-silicon/post-silicon; does it require \ac{RTL} source code), \textbf{Target} (on which target architecture was the fuzzer tested), \textbf{Input generation} (how is the input generated), \textbf{Platform} (how is the target \ac{CPU} run; emulated \ac{CPU}, \ac{CPU} on \ac{FPGA}, bare-metal, or requiring \ac{OS}), \textbf{\ac{ISA}-Simulation} (does the fuzzer require an additional \ac{ISA} simulator), \textbf{Vulnerability detection} (which method is used to detect bugs), \textbf{Microarchitectural feedback} (does it use \uarch feedback), and \textbf{Fuzzing input restriction} (is the fuzzing input generation restricted before running it on the target).

In the following, we highlight the differences between \ufuzz\ and the three previous post-silicon fuzzers. Osiris~\cite{osiris} is a post-silicon \xx\ fuzzer that detects timing-based side-channel vulnerabilities. It does so by measuring the duration instruction triplets take to execute. \ufuzz, in contrast, has to goal to find architectural and \uarch \ac{CPU} bugs by analyzing the \ac{CPU} state. SiliFuzz~\cite{silifuzz} initially fuzzes \ac{CPU} \ac{ISA} simulators using software coverage feedback to generate a test corpus and collect expected behavioral traces for each test case. Subsequently, the framework executes the generated corpus on the target \ac{CPU} and validates architectural register states against the simulator-derived expected values.
\ufuzz\ does not require an \ac{ISA} simulator and runs on the bare-metal \ac{CPU}, hence it does not restrict the input generation to ''non-destructive'' \xx\ instruction sequences. Like \ufuzz, RISCVuzz~\cite{RISCVuzz} does not use an \ac{ISA} Simulator, but it restricts the instruction generation to not include, e.g., \WRMSR\ equivalent instructions. Further, it targets the \riscv\ architecture and runs on top of an \ac{OS}.

\ufuzz\ is the first post-silicon \ac{CPU} fuzzer that leverages hypervisor environments to not restrict fuzzing input generation to ''non-destructive'' instructions. It runs on the bare-metal \ac{CPU} without an \ac{OS}. Further, it is the first \ac{CPU} fuzzer using \uarch state to enhance the fuzzing feedback, leaning towards the concept of pre-silicon fuzzers that modify the behavior of the target \ac{CPU}.

\section{Conclusion}
\label{sec:conclusion}
In this paper, we presented \ufuzz, the first post-silicon fuzzer for x86 \acp{CPU} that leverages \ucode-level feedback to systematically explore internal \uarch behavior. By introducing \ucode coverage as a novel guidance signal, \ufuzz opens a new direction for hardware fuzzing beyond architectural observability. Our system integrates a lightweight, hypervisor-based execution environment to ensure isolated, deterministic test runs on real silicon and introduces a serialization oracle to detect vulnerabilities without relying on formal specifications.

We address and formalize the core challenges of post-silicon \xx\ fuzzing 
\emph{\uarch invisibility}, \emph{absence of bug detection oracle}, \emph{non-deterministic execution}, and \emph{fault containment}. 
We demonstrate how \ufuzz overcomes these challenges through a principled design. Our evaluation shows that \ufuzz can uncover \findingcnt significant findings, including two previously unknown \ucode-level speculative-execution vulnerabilities (\textit{F2} and \textit{F3}) and automatically rediscover $\mu$Spectre vulnerabilities (\textit{F1}). \ufuzz achieved \totalcovp coverage of all hookable \ucode paths, setting a new baseline for introspective fuzzing on proprietary \acp{CPU}. Along the way, we introduce optimized instrumentation strategies that reduce instrumentation overhead by \overheadimprove compared to the baseline. Together, these contributions establish \ufuzz as a practical, efficient, and powerful framework for uncovering security-critical vulnerabilities in modern, closed-source processors.

\section*{Acknowledgment}
Our research work was partially funded by Intel’s Scalable Assurance Program, DFG-SFB 1119-236615297, the European Union under Horizon Europe Programme-Grant Agreement 101070537-CrossCon, NSF-DFG-Grant 538883423, and the European Research Council under the ERC Programme-Grant 101055025-HYDRANOS. This work does not in any way constitute an Intel endorsement of a product or supplier. Any opinions, findings, conclusions, or recommendations expressed herein are those of the authors and do not necessarily reflect those of Intel, the European Union, or the European Research Council. We would like to thank Jason Fung, Matthias Schunter, and Nassim Corteggiani from Intel for their valuable feedback.

\section*{Ethics Considerations}
This research explores the security properties of commercial x86 \acp{CPU} through post-silicon fuzzing and \ucode-level introspection. While our work targets undocumented processor internals, all experiments were conducted on isolated, non-networked hardware under tightly controlled conditions. We did not engage with production systems, user data, or shared infrastructure at any point during our evaluation.

In accordance with responsible disclosure protocols, we provided comprehensive documentation of our findings to Intel Corporation via \texttt{secure@intel.com}, 
including detailed descriptions of microcode-level anomalies, behavioral inconsistencies, potential speculative execution leakage paths, complete attack demonstrations, and a draft of our paper. Intel's security team acknowledged our research and validated the reported attack scenarios. Their assessment concluded that current mitigation strategies sufficiently address the identified vulnerabilities, resulting in their decision not to assign CVE designations to these discoveries.

No human subjects or personal data were involved; therefore, no institutional ethics review was required. However, we followed the principles outlined in the Menlo Report~\cite{menloreport}, emphasizing respect for persons, beneficence, and responsible stewardship of research outcomes. We aim to advance understanding of \ac{CPU} security while minimizing harm and supporting industry efforts to improve processor trustworthiness.

\bibliographystyle{IEEEtran}
\bibliography{research.bib}

\clearpage
\appendices

\section{Microcode Patching and Feedback}\label{alg:microcode-coverage-patch}

\begin{algorithm}
\caption{Instrumentation Logic for Microcode Hooking and Coverage Collection pseudo-code}

\begin{algorithmic}[1]
\State \textbf{entry\_}$i$:
\State \quad \texttt{JmpImm}(\texttt{hook\_handler\_$i$\_even}) 
\State \quad \texttt{JmpImm}(\texttt{hook\_handler\_$i$\_odd})
\vspace{0.5em}

\State \textbf{hook\_handler\_}$i$\_\textbf{even}:
\State \quad \textsc{StagingBuf}[$0xba00$, $0xbb00$] $\gets$ \texttt{R10}, \texttt{R14}
\State \quad \texttt{R10} $\gets 2 \cdot i + 0$
\State \quad \texttt{R14} $\gets$ \texttt{exit\_$i$\_even}
\State \quad \texttt{JmpImm}(\texttt{handler})
\vspace{0.5em}

\State \textbf{hook\_handler\_}$i$\_\textbf{odd}:
\State \quad \textsc{StagingBuf}[$0xba00$, $0xbb00$] $\gets$ \texttt{R10}, \texttt{R14}
\State \quad \texttt{R10} $\gets 2 \cdot i + 1$
\State \quad \texttt{R14} $\gets$ \texttt{exit\_$i$\_odd}
\State \quad \texttt{JmpImm}(\texttt{handler})
\vspace{0.5em}

\State \textbf{handler}:
\State \quad Save additional registers to \textsc{StagingBuf}
\State \quad \textsc{cov\_idx} $\gets$ $R10 \cdot 2 + 0x1000$
\State \quad \textsc{RAM}[\textsc{cov\_idx}] $\gets$ \textsc{RAM}[\textsc{cov\_idx}] $+ 1$
\State \quad \textsc{RAM}[$R10 \cdot 8 + 0x1400$] $\gets$ \texttt{RIP}
\State \quad Collect any other state as needed
\State \quad Restore registers from \textsc{StagingBuf}
\State \quad \texttt{JmpReg}(\texttt{R14})
\vspace{0.5em}

\State \textbf{exit\_}$i$\_\textbf{even}:
\State \quad \texttt{R10}, \texttt{R14} $\gets$ \textsc{StagingBuf}[$0xba00$, $0xbb00$]
\State \quad Instruction $i_0^\text{even}$
\State \quad \texttt{JmpImm}($a_0^\text{even}$)
\vspace{0.5em}

\State \textbf{exit\_}$i$\_\textbf{odd}:
\State \quad \texttt{R10}, \texttt{R14} $\gets$ \textsc{StagingBuf}[$0xba00$, $0xbb00$]
\State \quad Instruction $i_0^\text{odd}$
\State \quad \texttt{JmpImm}($a_0^\text{odd}$)
\end{algorithmic}
\end{algorithm}

\section{Serialized oracle and unrolling misaligned jumps} \label{appendix:misaligned-jumps}

\begin{figure}
  \begin{center}
  {\includegraphics[width=\columnwidth]{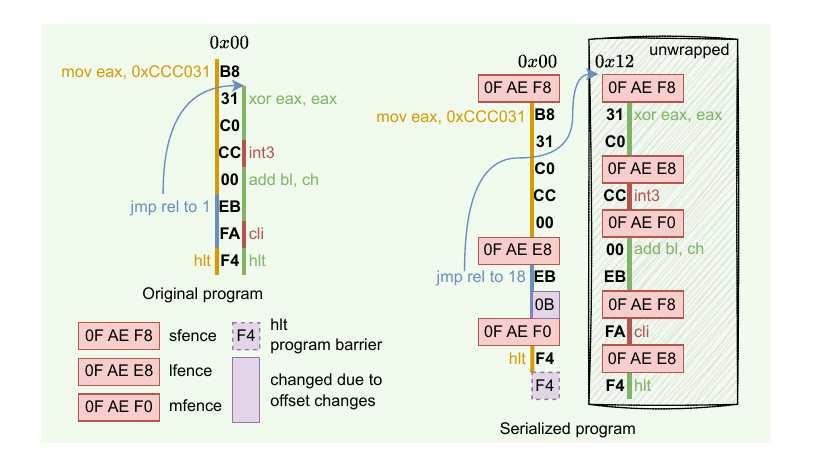}}
  \end{center}
  \caption{\label{fig:misaligned-jump} An exemplary \textit{serialization} of a program with an instruction-misaligned jump. The original program on the left contains a jump that jumps inside the payload of the \texttt{MOV} instruction. To \textit{serialize} the program, first, the instruction-misaligned instructions are unwrapped, then, both separate programs are serialized and joined using a \texttt{HLT} instruction.}
\end{figure}

During fuzzing, we encounter \texttt{jump/call/ret} target instructions that do not point to instruction-aligned boundaries. An example is visualized in Figure \ref{fig:misaligned-jump}: the \texttt{JMP} instruction on the left side jumps to the payload of the \texttt{MOV} instruction, continuing execution there.

When executing instructions misaligned, inserting fence instructions between regular instructions will change the execution of the program. In misaligned execution, instructions might be decoded that reach over several regular \xx\ instructions. By inserting fence instructions, the decoded instruction will, therefore, change, resulting in a different program output.

When encountering such a jump during \textit{serialization}, a new \textit{serialization} originating at the jump's target address (see Figure \ref{fig:misaligned-jump}) mus be started. We essentially suggest unwrapping the misaligned execution until the first illegal opcode and insert fence instructions between those instructions, like we would do for a regular instruction. This technique is detailed in Superset Assembly~\cite{binaryRewriting}. We propose joining the unwrapped and regular programs using, e.g., a \texttt{HLT} instruction, which will prevent the execution from continuing to the other program parts. Since all execution-flow changes are controlled, it can be ensured that the control flow will not cross the sub-program boundaries in an unintended fashion. This approach is only applicable for jumps with a target address known at \textit{serialization} time. 

During runtime address mapping, generally, the execution flow may be redirected to potentially any address of the original program, hence, the program must be unwrapped for each potential instruction offset. As a runtime optimization, the execution of a fuzzing input may be traced before \textit{serialization} using our hypervisor. Then, \textit{serialization} can be applied conditionally to relevant program parts only.

\section{Speculative fuzzing CPU lock-ups}\label{apdix:microcode_hang}

The complete dataset is available as a permanently archived artifact at \artifactDOI, accessible through the file \texttt{speculative\_fuzzing\_CPU\_lockups.csv}.

\begin{table*}[!htbp]
\centering
\caption{Comprehensive Catalog of Microcode Operations with Persistent Speculative Effects}
\label{tab:microcode_catalog_1}
\begin{tabular}{|c|c|l|}
\hline
\textbf{Instruction} & \textbf{Type} & \textbf{Disassembly} \\ \hline

\texttt{0x0e7500037033} & StableTimeout & \texttt{tmp7:= LDSTGBUF\_DSZ64\_ASZ16\_SC1(tmp3)} \\ \hline
\texttt{0x2e750003103a} & Unstable & \texttt{tmp1:= LDSTGBUF\_DSZ64\_ASZ16\_SC1(tmp10) !m2} \\ \hline
\texttt{0x0822c6df2232} & StableTimeout & \texttt{tmp2:= MOVETOCREG\_AND\_DSZ64(tmp2, 0x00000003, 0x7c6) !m0} \\ \hline
\texttt{0x0a62019c02f0} & Unstable & \texttt{MOVETOCREG\_BTR\_DSZ64(tmp0, 0x0000000e, 0x701) !m0} \\ \hline
\texttt{0x0a62019c02fb} & StableTimeout & \texttt{MOVETOCREG\_BTR\_DSZ64(tmp11, 0x0000000e, 0x701) !m0} \\ \hline
\texttt{0x0a62019c02fd} & StableTimeout & \texttt{MOVETOCREG\_BTR\_DSZ64(tmp13, 0x0000000e, 0x701) !m0} \\ \hline
\texttt{0x0a621c8002f0} & StableTimeout & \texttt{MOVETOCREG\_BTR\_DSZ64(tmp0, 0x0000000e, 0x01c) !m0} \\ \hline
\texttt{0x0a628c5002b0} & Unstable & \texttt{MOVETOCREG\_BTR\_DSZ64(tmp0, 0x00000009, 0x48c)} \\ \hline
\texttt{0x0a62c3180271} & Unstable & \texttt{MOVETOCREG\_BTR\_DSZ64(tmp1, 0x00000004, 0x6c3)} \\ \hline
\texttt{0x0a62c31802d4} & StableTimeout & \texttt{MOVETOCREG\_BTR\_DSZ64(tmpv0, 0x0000000c, 0x6c3)} \\ \hline
\texttt{0x0a62fe1c033a} & StableTimeout & \texttt{MOVETOCREG\_BTR\_DSZ64(tmp10, 0x00000010, 0x7fe)} \\ \hline
\texttt{0x0a62fe5c033a} & StableTimeout & \texttt{MOVETOCREG\_BTR\_DSZ64(tmp10, 0x00000011, 0x7fe)} \\ \hline
\texttt{0x0a62fe9c02b5} & StableTimeout & \texttt{MOVETOCREG\_BTR\_DSZ64(tmp5, 0x0000000a, 0x7fe) !m0} \\ \hline
\texttt{0x1a62cd880330} & StableTimeout & \texttt{MOVETOCREG\_BTR\_DSZ64(tmp0, 0x00000012, 0x2cd) !m0,m1} \\ \hline
\texttt{0x1a62cd880332} & StableTimeout & \texttt{MOVETOCREG\_BTR\_DSZ64(tmp2, 0x00000012, 0x2cd) !m0,m1} \\ \hline
\texttt{0x0962015c03b2} & StableTimeout & \texttt{MOVETOCREG\_BTS\_DSZ64(tmp2, 0x00000019, 0x701)} \\ \hline
\texttt{0x0962019c02ff} & StableTimeout & \texttt{MOVETOCREG\_BTS\_DSZ64(tmp15, 0x0000000e, 0x701) !m0} \\ \hline
\texttt{0x096204440370} & Unstable & \texttt{MOVETOCREG\_BTS\_DSZ64(tmp0, 0x00000015, 0x104)} \\ \hline
\texttt{0x096205040230} & Unstable & \texttt{MOVETOCREG\_BTS\_DSZ64(tmp0, 0x105)} \\ \hline
\texttt{0x09621cd747f4} & StableTimeout & \texttt{tmp4:= MOVETOCREG\_BTS\_DSZ64(tmp4, 0x0000003f, 0x51c) !m0} \\ \hline
\texttt{0x09623b1b13f1} & Unstable & \texttt{tmp1:= MOVETOCREG\_BTS\_DSZ64(tmp1, 0x0000001c, 0x63b)} \\ \hline
\texttt{0x096269000233} & StableTimeout & \texttt{MOVETOCREG\_BTS\_DSZ64(tmp3, 0x069)} \\ \hline
\texttt{0x096275d402b0} & Unstable & \texttt{MOVETOCREG\_BTS\_DSZ64(tmp0, 0x0000000b, 0x575) !m0} \\ \hline
\texttt{0x0962c3180273} & StableTimeout & \texttt{MOVETOCREG\_BTS\_DSZ64(tmp3, 0x00000004, 0x6c3)} \\ \hline
\texttt{0x0962e11c0200} & StableTimeout & \texttt{MOVETOCREG\_BTS\_DSZ64( , 0x7e1)} \\ \hline
\texttt{0x0962fe1c033d} & StableTimeout & \texttt{MOVETOCREG\_BTS\_DSZ64(tmp13, 0x00000010, 0x7fe)} \\ \hline
\texttt{0x19628f0c02b7} & StableTimeout & \texttt{MOVETOCREG\_BTS\_DSZ64(tmp7, 0x00000008, 0x38f) !m1} \\ \hline
\texttt{0x1962c10c0300} & StableTimeout & \texttt{MOVETOCREG\_BTS\_DSZ64( , 0x00000010, 0x3c1) !m1} \\ \hline
\texttt{0x1962c2480271} & Unstable & \texttt{MOVETOCREG\_BTS\_DSZ64(tmp1, 0x00000005, 0x2c2) !m1} \\ \hline
\texttt{0x1962cdc80330} & Unstable & \texttt{MOVETOCREG\_BTS\_DSZ64(tmp0, 0x00000013, 0x2cd) !m0,m1} \\ \hline
\texttt{0x0042011c0232} & StableTimeout & \texttt{MOVETOCREG\_DSZ64(tmp2, 0x701)} \\ \hline
\texttt{0x0042011f0230} & Unstable & \texttt{tmp0:= MOVETOCREG\_DSZ64(tmp0, 0x701)} \\ \hline
\texttt{0x00420400023f} & StableTimeout & \texttt{MOVETOCREG\_DSZ64(tmp15, 0x004)} \\ \hline
\texttt{0x00421a000200} & StableTimeout & \texttt{MOVETOCREG\_DSZ64( , 0x00000000, 0x01a)} \\ \hline
\texttt{0x00421c000214} & Unstable & \texttt{MOVETOCREG\_DSZ64(tmpv0, 0x01c)} \\ \hline
\texttt{0x00421d000238} & StableTimeout & \texttt{MOVETOCREG\_DSZ64(tmp8, 0x01d)} \\ \hline
\texttt{0x004229140200} & StableTimeout & \texttt{MOVETOCREG\_DSZ64( , 0x00000000, 0x529)} \\ \hline
\texttt{0x004229140235} & Unstable & \texttt{MOVETOCREG\_DSZ64(tmp5, 0x529)} \\ \hline
\texttt{0x004229140237} & StableTimeout & \texttt{MOVETOCREG\_DSZ64(tmp7, 0x529)} \\ \hline
\texttt{0x004229140238} & Unstable & \texttt{MOVETOCREG\_DSZ64(tmp8, 0x529)} \\ \hline
\texttt{0x00422914023b} & StableTimeout & \texttt{MOVETOCREG\_DSZ64(tmp11, 0x529)} \\ \hline
\texttt{0x004267000230} & StableTimeout & \texttt{MOVETOCREG\_DSZ64(tmp0, 0x067)} \\ \hline
\texttt{0x004267000231} & StableTimeout & \texttt{MOVETOCREG\_DSZ64(tmp1, 0x067)} \\ \hline
\texttt{0x004267000234} & StableTimeout & \texttt{MOVETOCREG\_DSZ64(tmp4, 0x067)} \\ \hline
\texttt{0x004267000235} & StableTimeout & \texttt{MOVETOCREG\_DSZ64(tmp5, 0x067)} \\ \hline
\texttt{0x004267000236} & StableTimeout & \texttt{MOVETOCREG\_DSZ64(tmp6, 0x067)} \\ \hline
\texttt{0x004267000238} & StableTimeout & \texttt{MOVETOCREG\_DSZ64(tmp8, 0x067)} \\ \hline
\texttt{0x004267000239} & StableTimeout & \texttt{MOVETOCREG\_DSZ64(tmp9, 0x067)} \\ \hline
\texttt{0x00426700023a} & StableTimeout & \texttt{MOVETOCREG\_DSZ64(tmp10, 0x067)} \\ \hline
\texttt{0x00426700023b} & StableTimeout & \texttt{MOVETOCREG\_DSZ64(tmp11, 0x067)} \\ \hline
\texttt{0x00426700023e} & StableTimeout & \texttt{MOVETOCREG\_DSZ64(tmp14, 0x067)} \\ \hline
\texttt{0x004270000230} & StableTimeout & \texttt{MOVETOCREG\_DSZ64(tmp0, 0x070)} \\ \hline
\texttt{0x004270000232} & StableTimeout & \texttt{MOVETOCREG\_DSZ64(tmp2, 0x070)} \\ \hline
\texttt{0x004277140230} & StableTimeout & \texttt{MOVETOCREG\_DSZ64(tmp0, 0x577)} \\ \hline
\texttt{0x00428e1c0230} & StableTimeout & \texttt{MOVETOCREG\_DSZ64(tmp0, 0x78e)} \\ \hline
\texttt{0x00428e1c0231} & StableTimeout & \texttt{MOVETOCREG\_DSZ64(tmp1, 0x78e)} \\ \hline
\texttt{0x00428e1c0232} & StableTimeout & \texttt{MOVETOCREG\_DSZ64(tmp2, 0x78e)} \\ \hline
\texttt{0x00428e1c0234} & StableTimeout & \texttt{MOVETOCREG\_DSZ64(tmp4, 0x78e)} \\ \hline
\texttt{0x00428e1c0239} & StableTimeout & \texttt{MOVETOCREG\_DSZ64(tmp9, 0x78e)} \\ \hline
\texttt{0x00428e1c023a} & StableTimeout & \texttt{MOVETOCREG\_DSZ64(tmp10, 0x78e)} \\ \hline
\texttt{0x00428e1c023b} & StableTimeout & \texttt{MOVETOCREG\_DSZ64(tmp11, 0x78e)} \\ \hline
\texttt{0x00429e1c0233} & StableTimeout & \texttt{MOVETOCREG\_DSZ64(tmp3, 0x79e)} \\ \hline
\end{tabular}
\end{table*}

\clearpage

\lstset{
   basicstyle=\ttfamily\footnotesize,
   breaklines=true,
   frame=single,
   backgroundcolor=\color{gray!10},
   commentstyle=\color{green!50!black},
   keywordstyle=\color{blue},
   stringstyle=\color{red},
   showstringspaces=false,
   tabsize=2,
   numbers=none,
   captionpos=b
}

\section{Artifact Appendix}\label{appendix:artifact}

This appendix assists users and future researchers in utilizing the \ourname artifact to reproduce the results presented in our paper, including fuzzing campaigns, coverage analysis, and proof-of-concept demonstrations for reported findings. Due to the specialized hardware requirements for setting up the \ourname framework and the complexity of configuration (particularly red-unlock \textit{Intel} \ac{CPU} microcode), we are applying only for the availability badge, following NDSS guidelines.

\ourname is an x86 \ac{CPU} fuzzer that leverages microcode coverage as feedback to guide fuzzing campaigns. The \ourname framework provides: (1) lightweight microcode instrumentation, (2) a minimal hypervisor for executing fuzzing inputs in isolated environments, (3) a serialization oracle for vulnerability detection, (4) core x86 \ac{CPU} fuzzing components including input generation, mutation engine, feedback collection, and vulnerability detection, (5) specialized capabilities for microcode-level fuzzing to detect speculative execution vulnerabilities, (6) coverage evaluation comparing microcode-guided fuzzing against baseline approaches, and (7) detection of seven findings.

The artifact includes source code for all \ourname components and experiments, along with comprehensive documentation for future research applications.

This appendix is organized as follows: Section~\ref{hwswconfreq} describes artifact access, hardware/software requirements, configurations, and benchmarks. Section~\ref{installconfig} provides high-level installation and configuration steps for environment preparation. Finally, Section~\ref{expworkflow} details the experimental workflow and instructions for reproducing each experiment described in the paper.

\subsection{Description \& Requirements}
\label{hwswconfreq}
The \ourname artifact has specific hardware, software, and configuration requirements. Each category is detailed below:

\subsubsection{How to access}
The complete source code for the \ourname{} framework is open-sourced at \github and hosted at the Zenodo permanent archival repository at \artifactDOI. These repositories contains source code and detailed documentation to reproduce all experimental results.

\subsubsection{Hardware dependencies}
The minimal hardware setup for running \ourname requires two Raspberry Pi 4 devices: (1) one serves as the fuzzer controller for generating test cases, collecting reports, and monitoring the target system, and (2) the second functions as a mass storage and virtual keyboard of the fuzzing agent.

Since \ourname targets x86 \acp{CPU} with microcode patch interface access capabilities, we selected the \textit{Intel Apollo Lake (Celeron, Goldmont)} N3350 processor (\texttt{CPUID[1].EAX=0x506ca}). This \ac{CPU} is integrated into the Gigabyte GB-BPCE-3350C board, which serves as our target platform. Given that the target \ac{CPU} may become unresponsive during fuzzing and require automatic restarts, we employ a breadboard with power switching capabilities. This setup enables remote power cycling of the Gigabyte GB-BPCE-3350C board by connecting to the board's power control pins through a relay circuit, controlled by the Raspberry Pi.

To facilitate data communication between the Raspberry Pi devices and the target \ac{CPU} board, we utilize a 5-port network switch for reliable network connectivity.

\subsubsection{Software dependencies}
The \ourname codebase is developed in the Rust programming language with the Cargo package manager for dependency management. We use NixOS as the \ac{OS} of both Pis with the full system configuration available in our artifact.

Additionally, the artifact requires the \texttt{uasm.py} file from the CustomProcessingUnit project~\cite{CustomProcessingUnit}. \ourname utilizes this script for compiling microcode updates during the fuzzing process.

\subsubsection{Configuration}
The \textit{Intel Apollo Lake (Celeron, Goldmont)} N3350 target \ac{CPU} does not natively provide access for applying customized microcode patches. To enable custom microcode patching capabilities, the \ac{CPU} must first be "red-unlocked".
The red-unlocking process follows established procedures documented in prior research~\cite{intelptxpoc,alaoui2019ptxe,ermolov2017hack}. Upon successful red-unlocking, users gain access to undocumented instructions that enable \ac{CRBUS} access (\texttt{udbgrd} and \texttt{udbgwr} instructions), which are essential for microcode manipulation during fuzzing operations.

\subsubsection{Benchmarks}
None

\subsection{Artifact Installation \& Configuration}
\label{installconfig}

This section provides the high-level installation and configuration steps required to prepare the environment for artifact evaluation. For detailed instructions, troubleshooting guidance, and comprehensive documentation, please refer to the \texttt{README.md} file included in the artifact.

\noindent \textbf{Dependency Installation:} The \ourname project requires specific dependencies on both the Raspberry Pi controllers and the development environment:

\begin{lstlisting}[language=bash]
# Dependencies
sudo apt install python3 python3-click \
     gcc-aarch64-linux-gnu build-essential git

# Rust Toolchain
curl --proto '=https' --tlsv1.2 -sSf https://sh.rustup.rs | sh -s -- --default-toolchain none -y

rustup install nightly-2025-05-30
rustup target add x86_64-unknown-uefi
rustup target add aarch64-unknown-linux-gnu
rustup target add x86_64-unknown-linux-gnu
rustup default nightly-2025-05-30


\end{lstlisting}

\noindent \textbf{CustomProcessingUnit~\cite{CustomProcessingUnit} Setup:} Download the CustomProcessingUnit project and place it in the parent directory of \ourname, or set the \texttt{UASM} environment variable to point to the \texttt{uasm.py} file location. Then apply the provided patch: 
\begin{lstlisting}[language=bash]
git apply uasm.py.patch
\end{lstlisting}

\noindent \textbf{Raspberry Pi Image Creation:}
Install Nix package manager following the official documentation, then build the SD card images for the Raspberry Pi devices:

\begin{lstlisting}[language=bash]
cd nix
# Configure SSH keys and IP addresses in configuration files
nix build .#images.master
nix build .#images.node
\end{lstlisting}
For subsequent deployments, use: \texttt{nix run}

\noindent \textbf{Target Device Setup:}
Deploy the \ac{UEFI} fuzzing application to the target \ac{CPU} board:

\begin{lstlisting}[language=bash]
HOST_NODE="<instrumentor_ip>" cargo xtask put-remote \
  --remote-ip <controller_address> \
  --source-ip <agent_address> \
  --netmask <network_mask> \
  --port <udp_port> \
  --startup <app_name>
\end{lstlisting}
Depending on the target fuzzing scenario, use \texttt{spec\_fuzz} (speculative microcode fuzzing) or \texttt{fuzzer\_device} (x86 instruction fuzzing) instead of \texttt{<app\_name>}.

\subsection{Experiment Workflow}
\label{expworkflow}
Start the fuzzer master and execute experiments using the provided command-line interface:
\begin{lstlisting}[language=bash]
fuzzer_master --help  # Available options/commands
\end{lstlisting}
Each of the following experiments includes built-in help documentation that allows you to specify parameters. For example:
\begin{lstlisting}[language=bash]
fuzz_master afl --help

# Executes the main fuzzing loop with AFL 
# mutations == Requires the `fuzzer_device` app 
# running on the agent ==

# Usage: fuzz_master afl [OPTIONS]

# Options:
#  -s, --solutions <SOLUTIONS>          
#  -c, --corpus <CORPUS>                
#  -a, --afl-corpus <AFL_CORPUS>        
#  -t, --timeout-hours <TIMEOUT_HOURS>  
#  -d, --disable-feedback               
#  -p, --printable-input-generation     
#  -h, --help                           
\end{lstlisting}

\noindent \textbf{Genetic Algorithm Fuzzing:} This mode performs coverage-guided fuzzing using a genetic mutation algorithm to evolve test cases based on microcode coverage feedback.
\begin{lstlisting}[language=bash]
fuzzer_master --database genetic_results.json \
  --instrumentor http://10.83.3.198:8000 \
  --agent 10.83.3.6:4444 \
  genetic
\end{lstlisting}

\noindent \textbf{AFL-based Fuzzing Campaign:}
This mode executes the main fuzzing loop using AFL (American Fuzzy Lop) mutation strategies with microcode coverage guidance for comprehensive x86 instruction fuzzing.
\begin{lstlisting}[language=bash]
fuzzer_master --database afl_campaign.json \
  --instrumentor http://10.83.3.198:8000 \
  --agent 10.83.3.6:4444 \
  afl
\end{lstlisting}

\noindent \textbf{Speculative Microcode Fuzzing:}
This mode performs specialized fuzzing targeting speculative execution vulnerabilities by manipulating microcode patches and monitoring speculative execution behavior.
\begin{lstlisting}[language=bash]
fuzzer_master --database spec_fuzzing.json \
  --instrumentor http://10.83.3.198:8000 \
  --agent 10.83.3.6:4444 \
  spec
\end{lstlisting}

\textbf{Note:} If you setup your agent and instrumentor on different IP addresses other than the default configuration, please modify the \texttt{--instrumentor} and \texttt{--agent} parameters accordingly in the above commands.

The fuzzing results will be stored to a database file (specifiable via the \textbf{--database} argument) (on fuzzer master) and can be viewed by running (on fuzzer master):
\begin{lstlisting}[language=bash]
fuzz_viewer database.json
\end{lstlisting}

\noindent \textbf{Reproducing Reported Findings}
The proof-of-concept implementations for the reported findings are located in the following directories within the artifact:

\begin{itemize}
\item \textit{Microcode Speculation (\texttt{CRBUS}-based):}
\begin{itemize}
    \item Location: \texttt{speculation\_ucode/src/main.rs}
    \item Description: Microcode speculation harness for \ac{CRBUS} interface
\end{itemize}

\item \textit{Microcode Speculation (\texttt{WRSGFLD}-based):}
\begin{itemize}
    \item Location: \texttt{fuzzer\_device/examples/test\_\\ucode\_speculation.rs}
    \item Description: Hypervisor speculation harness for \texttt{WRSGFLD} instruction
\end{itemize}

\item \textit{Speculative Window Termination:}
\begin{itemize}
    \item Location: \texttt{speculation\_x86/src/main.rs}
    \item Description: Terminating speculative execution window
\end{itemize}
\end{itemize}

\end{document}